\documentclass[]{aa}
\usepackage[dvips]{graphicx}
\usepackage{natbib}
\bibpunct{(}{)}{;}{a}{}{,}

\def\kms{km~s$^{-1}$}
\def\Teff{$T_{\rm eff}$}

\def\sch{Schwarzschild}

\def\p1{Paper~I}

\begin{document}


\title{Envelope tomography of long-period variable stars\thanks{Based on 
observations made at Observatoire de Haute Provence, operated by the Centre 
National de la Recherche Scientifique, France}}
\subtitle{III. Line-doubling frequency among Mira stars}
 
\author{Rodrigo Alvarez\inst{1}  
   \and Alain Jorissen\inst{1}\fnmsep\thanks{Research Associate, F.N.R.S. (Belgium)}
   \and Bertrand Plez\inst{2}
   \and Denis Gillet\inst{3}
   \and Andr\'e Fokin\inst{4}
   \and Maya Dedecker\inst{1}\fnmsep\thanks{F.R.I.A. Research Assistant (Belgium)}
}
\institute{Institut d'Astronomie et d'Astrophysique, 
           Universit\'e Libre de Bruxelles, 
           C.P.\,226, Boulevard du Triomphe,
           1050 Bruxelles, Belgium
           (ralvarez, ajorisse, dedecker@astro.ulb.ac.be)
\and
           GRAAL,
           Universit\'e Montpellier II,
           cc072,
           34095 Montpellier cedex 05, France
           (plez@graal.univ-montp2.fr)
\and
           Observatoire de Haute-Provence,
           04870 Saint-Michel l'Observatoire, France
           (gillet@obs-hp.fr)
\and
           Institute for Astronomy of the Russia Academy of Sciences,
           48 Pjatnitskaja,
           109017 Moscow, Russia
           (fokin@inasan.rssi.ru)
}

\offprints{A.\ Jorissen \email{ajorisse@astro.ulb.ac.be}}

\date{Received date / Accepted date}

\titlerunning{Envelope tomography of LPV stars. III}
\authorrunning{R.\ Alvarez et al.}

\abstract{
This paper presents statistics of the line-doubling phenomenon in a sample of
81 long-period variable (LPV) stars of various periods, spectral types and
brightness ranges. The set of observations consists of 315 
high-resolution optical spectra collected with the spectrograph ELODIE at the 
Haute-Provence Observatory, during 27 observing nights at one-month intervals and
spanning two years.  
When correlated with a mask mimicking a K0III spectrum, 54\% of the sample stars
clearly showed a double-peaked cross-correlation profile around maximum light,
reflecting double absorption lines. 
Several pieces of evidence are presented that point towards the double absorption
lines as being caused by the propagation of a shock wave through the photosphere. 
The observation of the Balmer lines appearing in emission 
around maximum light in these stars  
corroborates the presence of a shock wave. 
The observed velocity discontinuities, ranging between 10 and
25 km s$^{-1}$, are not correlated with the brightness ranges.
A comparison with the center-of-mass (COM) velocity obtained
from submm CO lines originating in the circumstellar envelope reveals that the
median velocity between the red and blue peaks is blueshifted with respect to
the COM velocity, as expected if the shock moves upwards. \\
The LPVs      
clearly exhibiting line-doubling around maximum light with the K0III mask 
appear to be the
most compact ones, the stellar radius being 
estimated from their effective temperatures (via the
spectral
type) and luminosities (via the period-luminosity relationship). 
It is not entirely clear whether or not this segregation between
  compact and extended LPVs is an
  artefact of the use of the K0III
  mask. Warmer masks (F0V and G2V) applied to the most extended
 and coolest LPVs yield asymmetric
  cross-correlation functions which suggest that line doubling is
  occurring in those stars as well. Although a firm conclusion on this
  point is hampered by the large correlation noise present in the
  CCFs of cool LPVs obtained with warm masks, the occurrence of line
  doubling in those stars is confirmed by the double
  CO $\Delta v = 3$ lines observed around 1.6 $\mu$m by Hinkle
  et al. (1984, ApJS 56, 1). 
Moreover, the H$\delta$ line in emission, which is another signature of the
presence of shocks, is observed as well 
in the most extended stars, 
although with a 
somewhat narrower profile. This is an indication that the shock is
weaker in extended than in compact LPVs, which may also contribute to
the difficulty of detecting line doubling in cool, extended LPVs. 
}

\maketitle

\section{Introduction}

Long-period variable stars (LPVs) are cool giant stars of low and 
intermediate masses, and the light variations associated with the
pulsation of their envelope have 
long periods (typically 300~d) and large amplitudes (up to 9 magnitudes 
peak-to-peak in the visual). LPVs are subdivided into Mira Ceti-type 
variables (Mira stars or Miras) and semi-regular variables (of the SRa or SRb 
subtypes), depending on the light 
curve amplitude, shape and regularity.

Apart from the brightness fluctuations, LPVs are characterized by striking 
spectral changes:
\begin{itemize}
\item appearance and disappearance of hydrogen and metallic emission lines
and splitting of many of the emission lines of the Balmer series into 
several components (e.g.\ Merrill 1921, 1923; Merrill \& Burwell 1930; 
Adams 1941; Joy 1947; Deutsch \& Merrill 1959; Gillet et al.\ 1983, 1985a, 
1985b, 1989; Ferlet \& Gillet 1984; Gillet 1988a, 1988b); 
\item absorption line shifts correlated with the lower excitation potential 
of the line (e.g.\ Adams 1941; Merrill 1947; Fujita 1951; Joy 1954);
\item doubling of several absorption lines around maximum light (e.g.\ 
Adams 1941; Merrill \& Greenstein 1958; Maehara 1968, 1971; Tsuji 1971; 
Hinkle 1978; Hinkle \& Barnes 1979a, 1979b; Hinkle et al.\ 1982, 1984, 1997;
Wallerstein 1985; Wallerstein et al.\ 1985; Barnbaum 1992; 
Hinkle \& Barnbaum 1996).
\end{itemize}

Merrill (1955) was the first to suggest that the spectral changes observed 
in LPVs may be explained by the presence of a shock wave moving outward. 
This idea has been extensively investigated since then by a number of 
authors: e.g.\ Gorbatskii (1957, 1961); Willson (1972, 1976); Hinkle
(1978); Willson \& 
Hill (1979); Hill \& Willson (1979); Wood (1979); Willson et al.\ (1982); 
Gillet \& Lafon (1983, 1984, 1990); Fox et al.\ (1984); Fox \& Wood (1985); 
Bessell et al.\ (1996); Fadeyev \& Gillet (2000, 2001).

Nevertheless, due to the lack of a complete and self-consistent model 
describing the pulsation and its effect on the spectrum, most of the
questions raised by the spectral peculiarities of LPVs remain
unanswered so far. For instance, the origin of the emission lines is still 
debated: although most authors believe they are formed in the hot wake
of the shock (Gillet 1988a; de la Reza 1986 and references therein), others 
reject the shock wave scenario and invoke instead purely non-LTE 
radiative processes (Magnan \& de Laverny 1997).

The velocity of the shock front is also a matter of debate: large 
shock-wave velocity discontinuities (of the order of 60~\kms) are indeed
required to 
photodissociate H$_2$ molecules and photoionize hydrogen atoms in order to 
subsequently produce by recombination the observed Balmer emission lines 
(Gillet et al.\ 1989). Velocity discontinuities of that order are indeed
observed for  fluorescent lines (Willson 1976), although 
double absorption lines yield much lower 
velocity discontinuities (of the order 
of 20--30~\kms; e.g.\ Adams 1941; 
Merrill \& Greenstein 1958; Maehara 1968; Tsuji 1971; Hinkle 1978; 
Hill \& Willson 1979; 
Hinkle et al.\ 1997 and references therein). The lower velocity discontinuities 
derived from absorption lines are not necessarily incompatible with the 
theoretical requirement, since the velocities of the blue and red components
suffer from a strong averaging effect due to the large extension of the region
where they form.

The line-doubling phenomenon is source of conflicting theories. Some 
important progress towards its understanding has
however been made recently thanks to a dedicated monitoring of Mira 
variables (Alvarez et al.\ 2000a, hereafter \p1), which confirmed that the 
line-doubling phenomenon is caused by a shock wave propagating in the 
photosphere. It was shown that the temporal evolution of the red and blue 
peaks of the double absorption lines of the Mira variable RT\,Cyg is 
consistent with the so-called ``\sch\ scenario''. This scenario, originally 
presented by \sch\ (1952) for W\,Vir Cepheids, relates the 
evolution of the line profile to the progression of a shock wave in the 
atmosphere. Alternative models accounting for line doubling without 
resorting to differential atmospheric motions (Karp 1975; Gillet et al.\ 
1985a) can thus be definitively rejected.

It was shown in \p1\ that some LPVs (e.g., X\,Oph) do {\it not}
(or, at least, not clearly) 
exhibit line doubling around maximum light. This paper further 
investigates the questions raised by this result: what is the frequency of
Mira variables exhibiting line-doubling around maximum light 
(Sect.~\ref{Sect:statistics}) and what are their distinctive properties 
(Sect.~\ref{Sect:schwarz})? To address these questions, a large sample of 
LPVs (mostly Mira variables) of various spectral types was
monitored during 2 years (Sect.~\ref{Sect:sample_obs}).

\section{Sample and observations}
\label{Sect:sample_obs}

This section describes the two-year-long monitoring of a large sample of LPV 
stars performed with ELODIE at the Observatoire de Haute-Provence (France) 
at a frequency of about one night per month, weather-permitting.

\subsection{The spectrovelocimeter ELODIE}
\label{Sect:ELODIE}

The fibre-fed echelle spectrograph ELODIE (Baranne et al.\ 1996) is mounted 
on the 1,93-m telescope of the Observatoire de Haute-Provence (France). This 
instrument is designed to perform very accurate radial-velocity measurements 
by cross-correlating the stellar spectrum with numerical masks. In one 
exposure an echelle spectrum at a resolution of 42\,000 ranging from 
3906~\AA\ to 6811~\AA\ is recorded. Then, an automatic on-line data 
processing extracts a 2-D spectrum (67 orders $\times$ 1044 pixels) from the
echelle spectrum. We developed our own cross-correlation procedure, following 
the prescriptions of Baranne et al.\ (1996). We have thus the possibility to 
use our own numerical templates and to perform different kinds of tests in 
the computation of the cross-correlation function (CCF). The CCF is computed 
directly from the 2-D spectrum. First, it is wavelength-calibrated, then 
corrected from the value of the Earth barycentric velocity towards the star. 
To preserve the accuracy, the CCF is computed without either rebinning or 
merging the orders. Only orders having an average signal-to-noise
(S/N) ratio 
greater than 2.0 are used.

Two templates have been systematically used in a first step to compute the 
CCFs of the whole sample: 
(i) the default K0\,III mask provided in the ELODIE reduction software 
(Baranne et al.\ 1996);
(ii) a M4\,V mask constructed by Delfosse et al.\ (1999) from an ELODIE 
spectrum of Barnard's star (Gl~699) applying the method of Baranne et 
al.\ (1979). 
Although the K0III template may in fact seem inadequate to study much
cooler LPVs, it was argued in Sect.~2.2 of Alvarez et al. (2001;
hereafter Paper~II) that `warm' masks like the K0III one are more
prone to detect line doubling (see also Sects.~\ref{Sect:CCF} and
4.3.2.1). On the other hand, it has been checked {\it a posteriori}
(see the discussion in Sect.~4.3.2.1 in relation with Fig.~\ref{Fig:RPsc})
that the M4V mask yields CCFs that are identical to those
  obtained with a mask mimicking the spectrum of the M4.5III star
  HD~123657 (see Table~\ref{Tab:sample1}).

\subsection{The sample}

The sample is composed of:
(i) 6 circumpolar Miras (SU\,Cam, S\,UMi, RT\,Dra, AX\,Cep, S\,Cep, RY\,Cep)
that can be observed all along the pulsational cycle;
(ii) 76 LPVs (Mira and semi-regular variables) of different periods, 
chemical compositions, spectral types and brightness ranges, observed at 
different phases;
(iii) 8 {\bf non-LPV} red giants;
(iv) 3 radial-velocity standard stars.
The entire sample (93 stars) is presented in Table~\ref{Tab:sample}: for 
each star are indicated its GCVS ({\it General Catalogue of Variable Stars}, 
Kholopov et al.\ 1988) name (or the HD number); its right ascension and  
declination (2000.0); its variability type, period, spectral type and
{\bf brightness range} 
as given by the GCVS; the number of observations $N$. The last column 
indicates whether the star is circumpolar ('circum.'), non-LPV 
('non-LPV') or radial-velocity standard ('r.v.\ st.').

\renewcommand{\baselinestretch}{0.7}
\begin{table*}
\caption[]{Sample stars}
\small{
\begin{flushleft}
\begin{tabular}{llllllrrl}
\hline\noalign{\smallskip}
Name & Right Asc. & Declination & Variability & Period & Spectral &
Brightness & $N$ & Comments\\
 & (2000.0) & (2000.0) & Type & (d) & Type & Range & & \\
\noalign{\smallskip}
\hline\noalign{\smallskip}
        SV And & 00:04:20 & +40:06:36 & Mira & 316.21 &      M5e-M7e & 
  7.7--14.3 &  2 &          \\
         T And & 00:22:23 & +26:59:46 & Mira & 280.76 &    M4e-M7.5e & 
  7.7--14.5 &  2 &          \\
         R And & 00:24:02 & +38:34:39 & Mira & 409.33 &  S3,5e-S8,8e & 
  5.8--14.9 &  1 &          \\
         Y Cep & 00:38:22 & +80:21:24 & Mira & 332.57 &    M5e-M8.2e & 
  8.1--16.0 &  1 &          \\
         U Psc & 01:22:58 & +12:41:54 & Mira & 173.10 &          M4e & 
 10.3--14.9 &  1 &          \\
         R Psc & 01:30:38 & +02:52:55 & Mira & 344.50 &      M3e-M6e & 
  7.0--14.8 &  5 &          \\
         Y And & 01:39:37 & +39:20:37 & Mira & 220.53 &    M3e-M4.5e & 
  8.2--15.1 &  5 &          \\
       $o$ Cet & 02:19:21 & $-$02:58:28 & Mira & 331.96 &      M5e-M9e & 
  2.0--10.1 &  3 &          \\
         R Cet & 02:26:02 & $-$00:10:42 & Mira & 166.24 &       M4e-M9 & 
  7.2--14.0 &  7 &          \\
         U Cet & 02:33:44 & $-$13:08:54 & Mira & 234.76 &      M2e-M6e & 
  6.8--13.4 &  1 &          \\
         R Tri & 02:37:02 & +34:15:54 & Mira &  266.9 &   M4IIIe-M8e & 
  5.4--12.6 &  3 &          \\
         U Ari & 03:11:03 & +14:47:58 & Mira & 371.13 &    M4e-M9.5e & 
  7.2--15.2 &  5 &          \\
        SS Cep & 03:49:30 & +80:19:20 &  SRb &   90.0 &        M5III & 
  8.0-- 9.1 &  3 &          \\
         R Tau & 04:28:18 & +10:09:44 & Mira & 320.90 &      M5e-M9e & 
  7.6--15.8 &  4 &          \\
         V Tau & 04:52:02 & +17:32:18 & Mira &  168.7 &    M0e-M4.5e & 
  8.5--14.6 &  4 &          \\
         R Aur & 05:17:18 & +53:35:11 & Mira & 457.51 &  M6.5e-M9.5e & 
  6.7--13.9 &  3 &          \\
         W Aur & 05:26:55 & +36:54:11 & Mira & 274.27 &      M3e-M8e & 
  8.0--15.3 &  2 &          \\
        RU Aur & 05:40:08 & +37:38:12 & Mira & 466.47 &      M7e-M9e & 
  9.0--16.0 &  3 &          \\
         S Cam & 05:41:02 & +68:47:55 &  SRa & 327.26 &   C7,3e(R8e) & 
  7.7--11.6 &  5 &          \\
  $\alpha$ Ori & 05:55:10 & +07:24:25 &  SRc &        &           M1 & 
 0.50       &  1 & non-LPV \\
         U Ori & 05:55:49 & +20:10:31 & Mira &  368.3 &    M6e-M9.5e & 
  4.8--13.0 &  4 &          \\
  $\delta$ Aur & 05:59:31 & +54:17:11 &      &        &           K0 & 
 3.71       &  1 & non-LPV \\
         X Aur & 06:12:13 & +50:13:41 & Mira & 162.79 &      M3e-M7e & 
  8.0--13.6 &  2 &          \\
     $\mu$ Gem & 06:22:57 & +22:30:54 &   Lb &        &           M3 & 
 2.97       &  1 & non-LPV \\
        SU Cam & 06:38:12 & +73:55:00 & Mira & 285.03 &           M5 & 
  8.9--12.6 &  8 &  circum. \\
         X Gem & 06:47:07 & +30:16:35 & Mira & 264.16 & M5e-M8e(Tc:) & 
  7.5--13.8 &  3 &          \\
         X Mon & 06:57:12 & $-$09:03:51 &  SRa & 155.80 &  M1eIII-M6ep & 
  6.8--10.2 &  1 &          \\
         R Gem & 07:07:21 & +22:42:13 & Mira & 369.91 &  S2.9e-S8.9e & 
  6.0--14.0 &  7 &          \\
         R CMi & 07:08:43 & +10:01:27 & Mira & 337.78 & C7,1Je(CSep) & 
 7.25--11.6 &  1 &          \\
         S CMi & 07:32:43 & +08:19:07 & Mira & 332.94 &      M6e-M8e & 
  6.6--13.2 &  5 &          \\
$\upsilon$ Gem & 07:35:55 & +26:53:50 &      &        &           M0 & 
 4.06       &  1 & non-LPV \\
        81 Gem & 07:46:08 & +18:30:39 &      &        &           K4 & 
 4.87       &  1 & non-LPV \\
         R Cnc & 08:16:34 & +11:43:35 & Mira &  361.6 &      M6e-M9e & 
 6.07--11.8 &  5 &          \\
         X UMa & 08:40:49 & +50:08:11 & Mira & 249.04 &      M3e-M4e & 
  8.1--14.8 &  4 &          \\
      HD 76830 & 08:59:11 & +18:08:09 &      &        &           M4 & 
 6.38       &  1 & non-LPV \\
        UZ Hya & 09:16:45 & $-$04:36:24 & Mira & 260.95 &          M4e & 
  8.8--14.5 &  5 &          \\
         R Leo & 09:47:33 & +11:25:46 & Mira & 309.95 &   M6e-M8IIIe & 
  4.4--11.3 &  3 &          \\
         S LMi & 09:53:43 & +34:55:32 & Mira & 233.83 &  M2.0e-M8.2e & 
  7.5--14.3 &  2 &          \\
         V Leo & 10:00:02 & +21:15:40 & Mira & 273.35 &          M5e & 
  8.4--14.6 &  4 &          \\
         R UMa & 10:44:39 & +68:46:33 & Mira & 301.62 &      M3e-M9e & 
  6.5--13.7 &  4 &          \\
        RU UMa & 11:41:40 & +38:28:30 & Mira & 252.46 &      M3e-M5e & 
  8.1--15.0 &  4 &          \\
         Y Vir & 12:33:52 & $-$04:25:18 & Mira & 218.43 &      M2e-M5e & 
  8.3--15.0 &  4 &          \\
         R Vir & 12:38:30 & +06:59:18 & Mira & 145.63 &  M3.5IIIe-M8 & 
  6.1--12.1 &  2 &          \\
        RS UMa & 12:38:57 & +58:29:03 & Mira & 258.97 &      M4e-M6e & 
  8.3--14.9 &  4 &          \\
         S UMa & 12:43:57 & +61:05:36 & Mira & 225.87 &  S0,9e-S5,9e & 
  7.1--12.7 &  2 &          \\
         U Vir & 12:51:06 & +05:33:12 & Mira & 206.64 &     M2e-M8e: & 
  7.4--13.5 &  2 &          \\
         V UMi & 13:38:41 & +74:18:37 &  SRb &   72.0 &      M5IIIab & 
  7.2-- 9.1 &  5 &          \\
        SY Vir & 13:58:38 & $-$04:34:35 & Mira & 236.65 &          M6: & 
  9.6--13.4 &  1 &          \\
     HD 123657 & 14:07:56 & +43:51:18 &   Lb &        &      M4.5III & 
 5.25       &  1 & non-LPV \\
  $\alpha$ Boo & 14:15:38 & +19:11:06 &      &        &      K1.5III & 
-0.04       &  1 & r.v.\ st \\
         R Boo & 14:37:12 & +26:44:12 & Mira &  223.4 &      M3e-M8e & 
  6.2--13.1 &  5 &          \\
         Y Lib & 15:11:41 & $-$06:00:43 & Mira & 275.70 &    M5e-M8.2e & 
  7.6--14.7 &  4 &          \\
        RT Boo & 15:17:15 & +36:21:34 & Mira & 273.86 &    M6.5e-M8e & 
  8.3--13.9 &  2 &          \\
         S Ser & 15:21:40 & +14:18:52 & Mira & 371.84 &      M5e-M6e & 
  7.0--14.1 &  1 &          \\
         S UMi & 15:29:35 & +78:38:00 & Mira &  331.0 &      M6e-M9e & 
  7.5--13.2 & 10 &  circum. \\
        ST Her & 15:50:47 & +48:29:00 &  SRb &  148.0 &           M6 & 
  8.8--10.3 &  1 &          \\
        RU Her & 16:10:15 & +25:04:14 & Mira & 484.83 &       M6e-M9 & 
  6.8--14.3 &  7 &          \\
        SS Oph & 16:57:52 & $-$02:45:42 & Mira & 180.64 &          M5e & 
  7.8--14.5 &  1 &          \\
        RV Her & 17:00:35 & +31:13:22 & Mira & 205.23 &          M2e & 
  9.0--15.5 &  1 &          \\
        SY Her & 17:01:29 & +22:28:40 & Mira & 116.91 &      M1e-M6e & 
  8.4--14.0 &  1 &          \\
         Z Oph & 17:19:32 & +01:30:52 & Mira &  348.7 &    K3ep-M7.5 & 
  7.6--14.0 & 10 &          \\
        RS Her & 17:21:42 & +22:55:16 & Mira & 219.70 &       M4e-M8 & 
  7.0--13.0 &  3 &          \\
        RU Oph & 17:32:53 & +09:25:24 & Mira & 202.29 &      M3e-M5e & 
  8.6--14.2 &  1 &          \\
   $\beta$ Oph & 17:43:27 & +04:34:03 &      &        &        K2III & 
 2.77       &  6 & r.v.\ st \\
         T Her & 18:09:06 & +31:01:16 & Mira & 164.98 &    M2,5e-M8e & 
  6.8--13.7 &  1 &          \\
        RY Oph & 18:16:37 & +03:41:34 & Mira & 150.41 &       M3e-M6 & 
  7.4--13.8 &  1 &          \\
        RT Dra & 18:19:26 & +72:40:50 & Mira & 279.41 &           M5 & 
  9.6--13.8 & 14 &  circum. \\
        SV Her & 18:26:23 & +25:01:35 & Mira & 238.99 &          M5e & 
  9.1--15.1 &  1 &          \\
         X Oph & 18:38:21 & +08:50:01 & Mira & 328.85 &      M5e-M9e & 
  5.9-- 9.2 &  4 &          \\
        WZ Lyr & 19:02:15 & +47:12:56 & Mira & 376.64 &          M9e & 
 10.6--15.0 &  2 &          \\
        RU Lyr & 19:12:21 & +41:18:12 & Mira & 371.84 &     M6e:-M8e & 
  9.5--15.9 &  1 &          \\
         W Aql & 19:15:23 & $-$07:02:50 & Mira & 490.43 &  S3,9e-S6,9e & 
  7.3--14.3 &  1 &          \\
        RT Cyg & 19:43:38 & +48:46:41 & Mira & 190.28 &  M2e-M8.8eIb & 
  6.0--13.1 &  7 &          \\
  $\gamma$ Aql & 19:46:14 & +10:36:43 &      &        &         K3II & 
 2.72       &  3 & r.v.\ st \\
    $\chi$ Cyg & 19:50:34 & +32:54:53 & Mira & 408.05 & S6,2e-S10,4e & 
  3.3--14.2 &  6 &          \\
\noalign{\smallskip}
\hline
\end{tabular}
\end{flushleft}}
\label{Tab:sample1}
\end{table*}
      
\addtocounter{table}{-1}
\begin{table*}
\caption[]{Sample stars: continued}
\small{
\begin{flushleft}
\begin{tabular}{llllllrrl}
\hline\noalign{\smallskip}
Name & Right Asc. & Declination & Variability & Period & Spectral &
Brightness & $N$ & Comments\\
 & (2000.0) & (2000.0) & Type & (d) & Type & Range & & \\
\noalign{\smallskip}
\hline\noalign{\smallskip}
         Z Cyg & 20:01:27 & +50:02:34 & Mira & 263.69 &      M5e-M9e & 
  7.1--14.7 &  1 &          \\
         S Aql & 20:11:37 & +15:37:13 &  SRa & 146.45 &    M3e-M5.5e & 
  8.9--12.8 &  1 &          \\
         Z Aql & 20:15:11 & $-$06:09:04 & Mira & 129.22 &          M3e & 
  8.2--14.8 &  1 &          \\
        WX Cyg & 20:18:34 & +37:26:54 & Mira & 410.45 & C8,2JLi(N3e) & 
  8.8--13.2 &  2 &          \\
         T Cep & 21:09:32 & +68:29:28 & Mira & 388.14 &  M5.5e-M8.8e & 
  5.2--11.3 &  6 &          \\
        RR Aqr & 21:15:01 & $-$02:53:43 & Mira & 182.45 &      M2e-M4e & 
  9.1--14.4 &  1 &          \\
         X Peg & 21:21:00 & +14:27:00 & Mira & 201.20 &      M2e-M5e & 
  8.8--14.4 &  1 &          \\
        SW Peg & 21:22:29 & +21:59:45 & Mira & 396.33 &          M4e & 
  8.0--14.0 &  2 &          \\
        AX Cep & 21:26:52 & +70:13:18 & Mira &  395.0 &         C(N) & 
  9.5--13.0 & 11 &  circum. \\
         S Cep & 21:35:13 & +78:37:28 & Mira & 486.84 &   C7,4e(N8e) & 
  7.4--12.9 & 12 &  circum. \\
        RV Peg & 22:25:38 & +30:28:22 & Mira & 396.80 &          M6e & 
  9.0--15.5 &  1 &          \\
        AR Cep & 22:51:33 & +85:02:47 &  SRb &        &        M4III & 
  7.0-- 7.9 &  2 & non-LPV \\
         R Peg & 23:06:39 & +10:32:35 & Mira &  378.1 &      M6e-M9e & 
  6.9--13.8 &  4 &          \\
         W Peg & 23:19:50 & +26:16:44 & Mira &  345.5 &      M6e-M8e & 
  7.6--13.0 &  5 &          \\
         S Peg & 23:20:33 & +08:55:08 & Mira & 319.22 &    M5e-M8.5e & 
  6.9--13.8 &  3 &          \\
        RY Cep & 23:21:14 & +78:57:31 & Mira & 149.06 &       Ke-M0e & 
  8.6--13.6 &  9 &  circum. \\
        ST And & 23:38:45 & +35:46:17 &  SRa & 328.34 &  C4,3e-C6,4e & 
  7.7--11.8 &  5 &          \\
         R Cas & 23:58:24 & +51:23:18 & Mira & 430.46 &     M6e-M10e & 
  4.7--13.5 &  4 &          \\
\noalign{\smallskip}
\hline
\end{tabular}
\end{flushleft}}
\label{Tab:sample}
\end{table*}

\renewcommand{\baselinestretch}{1.5}

\subsection{The observations}

A total of 27 observing nights on ELODIE were allocated between 1998, August 
and 2000, August (more or less one night per month). Approximatively one 
third of the total observing time was lost mainly because of bad atmospheric 
conditions, causing interruptions in the phase coverage. About 15--20 
stars were observed during each clear night, with typical exposure times of 
about 25~min 
yielding a S/N ratio per resolution element at 500~nm 
up to 150 for the brightest stars. For the faintest and reddest stars,
sometimes only the reddest orders recorded an usable signal. But the
power of a correlation technique is precisely that it does not
require high S/N spectra to deliver useful CCFs (see e.g.,
Queloz 1995). As indicated in Sect.~\ref{Sect:ELODIE},  
the CCF is computed using orders having an average S/N
ratio of at least 2.0. 

The log of observations is presented in Table~\ref{Tab:obs_log}: the first 
and second columns list the civil date and the corresponding Julian
date at midnight.
The third column gives the code number 
of the night: each observing night will be subsequently referred to by this 
number. Only the nights during which at least one star was observed 
are reported in Table~\ref{Tab:obs_log}.

A total of 315 spectra were collected, i.e.\ 3--4 spectra per star on average. 
Some (circumpolar) stars were observed as often as 12 times 
(S\,Cep), and the majority of variable stars were observed 2--3 times.

\begin{table}
\caption[]{Log of observations}
\begin{flushleft}
\begin{tabular}{lll}
\hline\noalign{\smallskip}
Date of      & Julian Date  & Night  \\
Observations & (2450000+)   & Number \\
\noalign{\smallskip}
\hline\noalign{\smallskip}
1998 Aug., 05--06 & 1031.5 &  N1 \\
1998 Sep., 03--04 & 1060.5 &  N2 \\ 
1998 Sep., 28--29 & 1085.5 &  N3 \\
1998 Oct., 06--07 & 1093.5 &  N4 \\
1998 Dec., 23--24 & 1171.5 &  N5 \\
1999 Jan., 26--27 & 1205.5 &  N6 \\
1999 Feb., 23--24 & 1233.5 &  N7 \\
1999 Apr., 22--23 & 1291.5 &  N9 \\
1999 May, 19--20  & 1318.5 & N10 \\
1999 Jul., 05--06 & 1365.5 & N12 \\
1999 Jul., 26--27 & 1386.5 & N13 \\
1999 Sep., 02--03 & 1424.5 & N14 \\
1999 Sep., 28--29 & 1450.5 & N15 \\
1999 Nov., 16--17 & 1499.5 & N16 \\
1999 Dec., 16--17 & 1529.5 & N17 \\
2000 Jan., 10--11 & 1554.5 & N18 \\
2000 Feb., 22--23 & 1597.5 & N19 \\  
2000 Feb., 23--24 & 1598.5 & N20 \\  
2000 Apr., 17--18 & 1652.5 & N22 \\
2000 May., 18--19 & 1683.5 & N23 \\
2000 Jun., 20--21 & 1716.5 & N24 \\
2000 Jul., 11--12 & 1737.5 & N25 \\
2000 Aug., 09--10 & 1766.5 & N26 \\
2000 Aug., 10--11 & 1767.5 & N27 \\  
\noalign{\smallskip}
\hline
\end{tabular}
\end{flushleft}
\label{Tab:obs_log}
\end{table}

\section{The variety of cross-correlation profiles and their frequency
of occurrence}
\label{Sect:CCF}

\begin{figure*}
  \resizebox{\hsize}{!}{\rotatebox{90}{\includegraphics{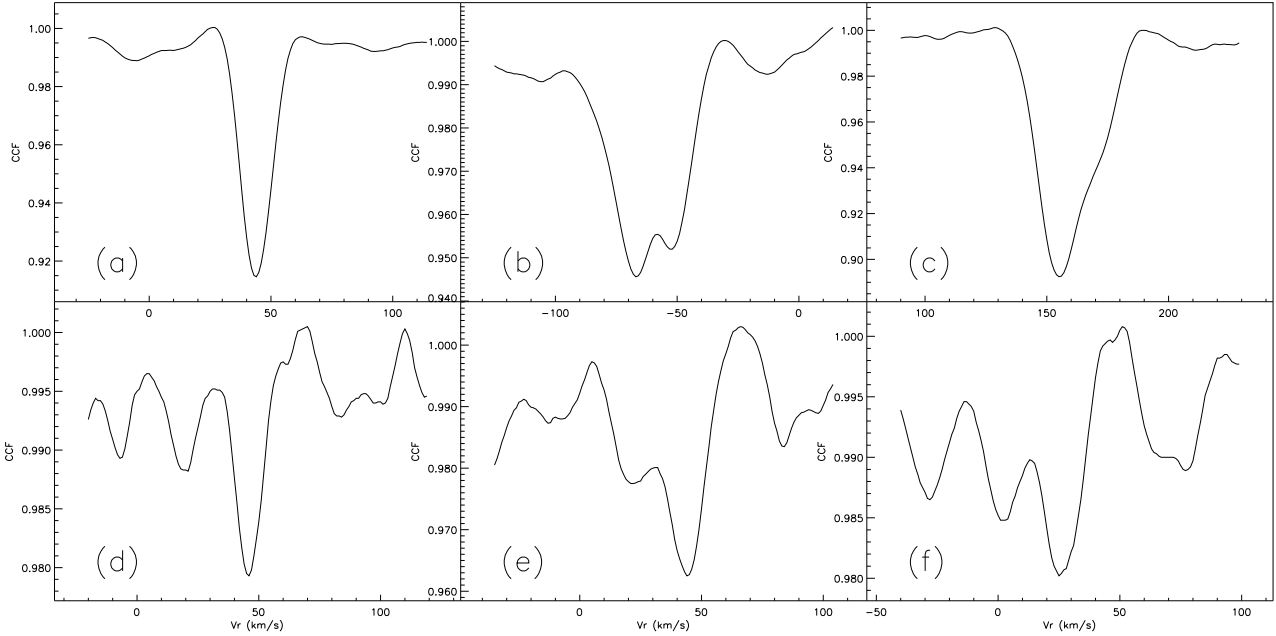}}}
  \caption[]{Examples of the different kinds of CCFs observed: 
  (a) {\it single peak}: R\,Cet, night N7, phase 0.96:, mask M4V; 
  (b) {\it double peak}: R\,Boo, night N7, phase 0.98, mask K0III;
  (c) {\it asymmetrical peak}: X\,Mon, night N22, phase 0.94, mask K0III;
  (d) {\it doubtful single peak}: R\,Cet, night N18, phase 0.89, mask K0III;
  (e) {\it doubtful double peak}: R\,Cet, night N4, phase 0.30, mask K0III;
  (f) {\it noisy profile}: R\,Cnc, night N3, phase 0.96, mask K0III}
  \label{Fig:types_ccf}
\end{figure*}

\begin{figure}
  \resizebox{\hsize}{!}{\includegraphics{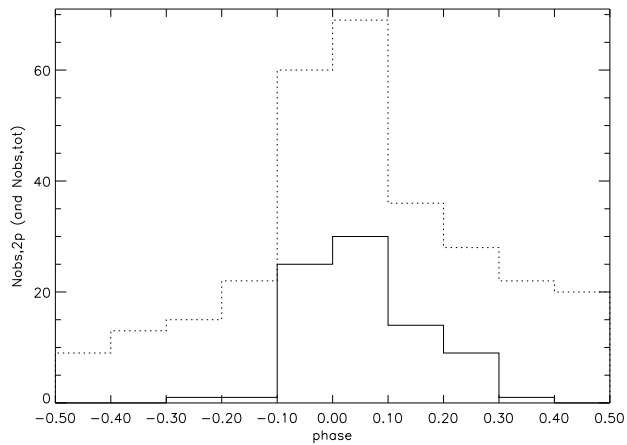}}
  \caption[]{Phase distributions (i) for the observations which  
  exhibit a double or an asymmetrical peak  with the K0-template (solid 
  line) and (ii) for the total set of observations (dashed line)}
  \label{Fig:h_2p_phi_K0}
\end{figure}

For each of the 315 spectra, we compute the CCFs
with the K0\,III and the M4\,V default templates. A large variety of CCFs 
are observed among LPVs, ranging from the classical single-peak profile to 
much more complex profiles (asymmetrical  peak, double peak, ``noisy'' 
profiles, etc.). We developed an automatic classification procedure which 
classifies the CCFs according to their shape and contrast into a finite 
number of archetypes, thus avoiding any subjectivity. 
Table~3 provides for each observation the CCF code for
the K0- and 
M4-templates, together with the night number, the Julian date, the phase 
and, when appropriate, the heliocentric radial velocities as derived by a 
single or double gaussian fit of the CCF.
The different kinds of CCFs observed are as 
follows:

\begin{itemize}

\item 
{\bf Single peak}: the most common profile observed is the classical single 
peak (Fig.~\ref{Fig:types_ccf}a). In this case, the minimum of the CCF is 
supposedly the radial velocity of the star, at least for the non-variable 
stars. For variable stars, the observed velocity might be perturbed 
by atmospheric motions and may thus not correspond to the center-of-mass
velocity (it is worth noting that these intrinsic radial-velocity variations
render the search for binaries among LPVs extremely difficult; see 
Sect.~\ref{Sect:binary}).

Single-peak profiles occur at all phases. 54\% of the LPV 
sample stars showed at least once this kind of profile with the 
K0-template (28\% of the total number of observations). These values amount
respectively to 89\% and 74\% with the M4-template. This cool template
thus yields single-peak,
well-contrasted CCFs much more often than the K0-template. These profiles 
are coded '1p' in Table~3
\footnote{only available in the electronic version of this paper; see also 
{\tt http://www-astro.ulb.ac.be/tomography.html}}.

\item
{\bf Double peak and asymmetrical peak}: the cross-correlation technique 
permits to very clearly reveal the doubling of the lines despite the severe 
crowding of the LPV spectra (Fig.~\ref{Fig:types_ccf}b). Sometimes, the 
profile only exhibits an asymmetry, as if the double peak was about to 
appear (or to disappear) (Fig.~\ref{Fig:types_ccf}c). 

A double or asymmetrical peak was observed for 50\% of the LPV 
stars (28\% of the total number of observations) with the K0-template. Most 
of them were not known in the literature to exhibit the line-doubling 
phenomenon. The doubling is essentially observed around maximum light 
(between phases $-$0.1 and 0.3), as can be seen on
Fig.~\ref{Fig:h_2p_phi_K0}. With the M4-template, only 23\% of sample stars 
(10\% of the total number of observations) showed these kinds of 
profiles (see Sect.~2.2 of Paper~II for a discussion of the physical 
origin of this difference between the K0- and the M4-templates), and the 
phases range from $-$0.2 to 0.6. These profiles are coded '2p' (double peak) 
or 'ap' (asymmetrical peak) in Table~3.

\item 
{\bf Doubtful single/double peak and noisy profile}: the profiles cannot 
always be unambiguously classified as single or double peak, because 
of the poor contrast and/or the complex shape of the CCFs. Such profiles 
can only be classified as ``doubtful single profiles'' or ``doubtful double 
profiles'' (Figs.~\ref{Fig:types_ccf}d and \ref{Fig:types_ccf}e, coded 
1p? and 2p?, respectively, in Table~3). In some 
cases, the contrast of the CCF is so low that the cross-correlation profile 
does not even yield any clear dip; no radial velocity can be derived from 
such noisy CCFs (Fig.~\ref{Fig:types_ccf}f), identified as ``noisy 
profiles'' (coded `np' in Table~3). 
The distinction between ``doubtful single/double-peak
profile'' and ``noisy profile'' is somewhat arbitrary 
 (compare Figs.~\ref{Fig:types_ccf}e and \ref{Fig:types_ccf}f) and depends 
upon the particular (and somewhat arbitrary) values adopted for the 
parameters involved in our automatic classification procedure. 
No attempt to 
derive radial velocities was made with these kinds of profiles. 

It must be noted that the general structure of the CCFs 
presented in Figs.~\ref{Fig:types_ccf}d and \ref{Fig:types_ccf}e is 
in fact strikingly similar, the only difference being the respective contrasts of 
the leftmost and second-to-leftmost peaks. Moreover, a given star (like
R Cet; see Table~\ref{Tab:ccf_continued} and Fig.~\ref{Fig:types_ccf})
may evolve from 1p? CCFs to 2p? CCFs and even to 2p at different phases of its light
cycle, thus suggesting that the 1p? and 2p? CCFs of the kind
displayed in Figs.~\ref{Fig:types_ccf}d and \ref{Fig:types_ccf}e  
may be intrinsically similar and should in fact be
classified together (that conclusion may in fact even be extended
to some 'np' profiles, like the one displayed in
Fig.~\ref{Fig:types_ccf}f whose general structure resembles that of
Figs.~\ref{Fig:types_ccf}d and \ref{Fig:types_ccf}e). 
However, in this first analysis of 
the statistics of the line-doubling phenomenon, it was decided to
adopt conservative classification criteria, at
the risk of excluding physically-sound data. 
In particular, the line-doubling statistics presented
in Figs.~\ref{Fig:h_2p_phi_K0}, \ref{Fig:h_2p_per_K0},
\ref{Fig:h_2p_spt_K0}, \ref{Fig:h_radius} and  \ref{Fig:h_radius_M4}
relies only on CCFs classified as `2p' or `ap' and excludes CCFs
classified as `2p?' and `np'.
The possible physical implications of this conservative
choice are discussed further in Sect.~\ref{Sect:statistics}.1.

Doubtful or noisy profiles were obtained at least once for 61\%
of the LPV sample stars (44\% of the total number of observations) with the 
K0-template. Although these profiles were found at all phases, they 
are preferentially observed around minimum light (when the star is fainter 
and cooler). With the M4-template, the above percentages turn to only 17\% 
of the stars (most of which are the S-type and C-type stars of the sample) 
and 16\% of the total number of observations. 
\end{itemize}

Figure~\ref{Fig:h_2p_phi_K0} presents the distribution in phase of the 
double-peak profiles as compared to the total number of observations. The
striking features exhibited by Fig.~\ref{Fig:h_2p_phi_K0} are (i) the very 
sharp rise in the fraction of double-peak profiles at phase $-$0.1, (ii) the 
total absence of double-peak profiles between phases 0.4 and 0.7, and (iii) 
the fraction of double-peak profiles remains almost constant ($\sim$40\%) 
between phases $-$0.1 and 0.3.

Another conclusion that can be drawn at this point concerns the interest of
the cross-correlation technique for dynamical studies: as already pointed 
out in \p1, the cross-correlation technique provides a powerful tool to 
extract double lines despite the severe crowding of the spectra of giant 
stars (see however the discussion of
  Sect.~\ref{Sect:statistics}.1 relative to late-type LPVs). 
The study of the spectral variations associated with the pulsation of 
LPV stars has no more to be restricted to the few clean near-IR spectral 
lines. The much richer visible spectrum is now accessible as well to these 
studies, opening great potentialities as illustrated for instance by the
tomographic technique described in Paper~II.

\section{The \sch\ scenario}
\label{Sect:schwarz}

The origin of the line-doubling phenomenon in LPVs has been a long-standing
problem (see \p1\ for a review). The most definitive way to prove that 
the line doubling is not the result of purely radiative processes but is
actually caused by the differential bulk motions in the atmosphere arising
from the propagation of a shock wave is to check whether or not the temporal 
sequence of the line doubling follows the \sch\ scenario.

\sch\ (1952) developed his scenario for W\,Vir variables, but 
in principle it can be generalized to any variable star where a shock 
wave propagates through the photosphere. According to this scenario, the 
temporal evolution of the intensity of the components of a double line 
should obey a precise sequence: the intensity of the blue component, formed
in deeper, ascending photospheric layers, should increase from zero to
maximum, while at the same time the red component, formed in the upper
infalling photospheric layers decreases from maximum to zero 
(Fig.~1 of \p1). 

Thanks to a two-month-long monitoring of two LPVs 
around maximum light, it was shown in \p1\ that (at least some) LPVs do
follow the \sch\ scenario. The present set of observations of many LPVs spread
over two
years enables us to answer the following questions in relation with the
line-doubling phenomenon: 
(i) at which phases does it occur?
(ii) does it repeat identically from one cycle to the next?
(iii) what is the fraction of LPV stars exhibiting the line doubling
phenomenon and what are their distinctive properties?

\subsection{Temporal evolution of the line doubling}
\label{Sect:temporal_evol}

Although it was shown in \p1\ that the line profile of the Mira RT\,Cyg
followed the \sch\ scenario around maximum light, there was no information
available about the evolution of the profile at later phases.

\begin{figure}
  \resizebox{\hsize}{!}{\includegraphics{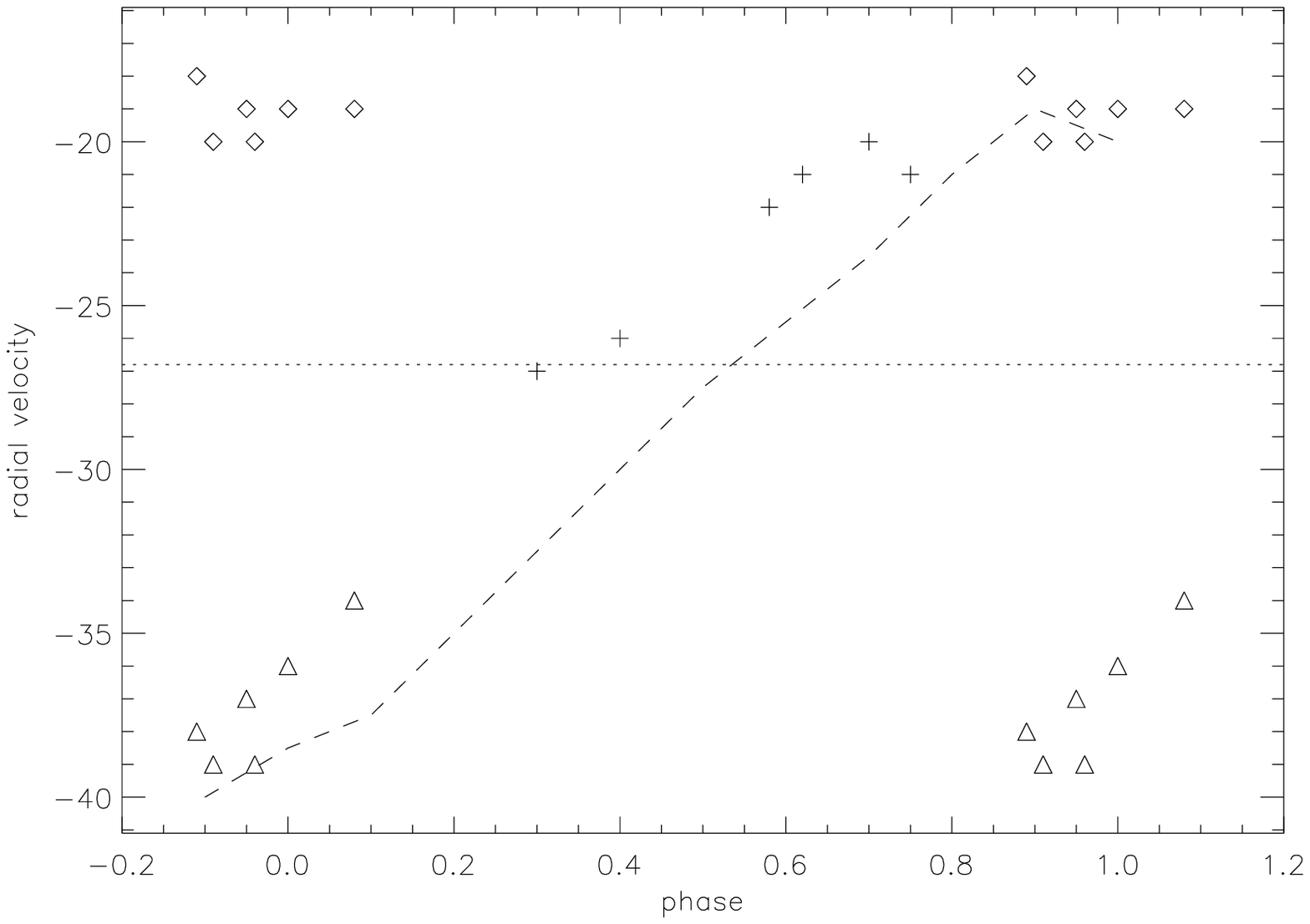}}
  \caption[]{Velocity variations of S\,Cep (cross: single component; 
  triangles: blue component; diamonds: red component). The dotted line 
  indicates the center-of-mass velocity (see
  Table~\ref{Tab:com_vr}). The velocities between phases 0.8 and 1.1
  are shown twice to illustrate the velocity behaviour through
  an entire cycle. The dashed line is an eye fit to the
  radial-velocity curve of S Cep obtained by Hinkle \& Barnbaum (1996) 
  from CN $\Delta v = -2$ red-system lines around 2 $\mu$m (their Fig.~1)}
  \label{Fig:scep_survey_K0_1}
\end{figure}

In the present sample, the circumpolar Mira S\,Cep (C7,4e; $P$=486.84~d 
according to the GCVS) has a good phase coverage and allows to draw the 
velocity curve over a full light cycle (Fig.~\ref{Fig:scep_survey_K0_1}): 
a typical S-shaped curve emerges that is very similar to the curves obtained
from the rotation-vibration CO or CN lines in the near-infrared (Hinkle et al.\
1997 and references therein). S\,Cep was already known to exhibit the
line-doubling phenomenon for those lines (Hinkle \& Barnbaum 1996;
Fig.~\ref{Fig:scep_survey_K0_1}).

\begin{figure*}
  \resizebox{\hsize}{!}{\rotatebox{90}{\includegraphics{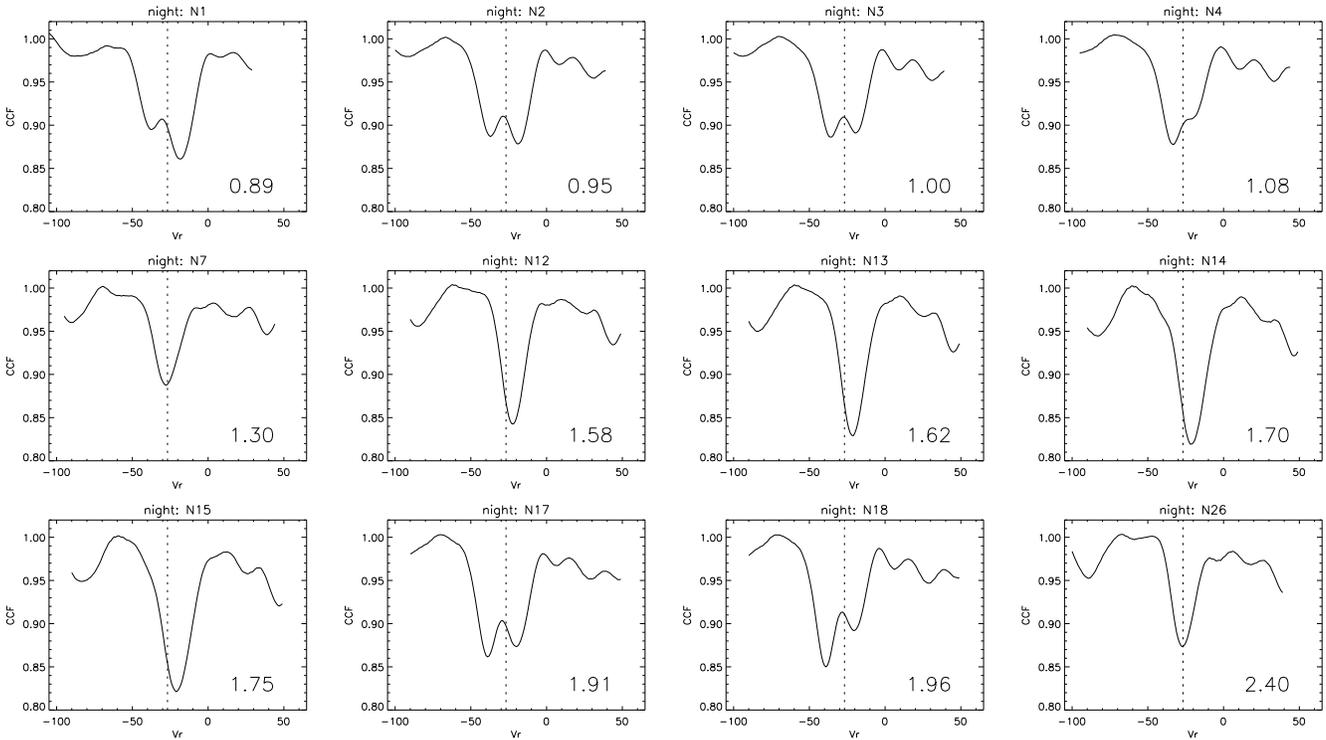}}}
  \caption[]{Sequence of CCFs for S\,Cep with the default K0\,III mask.
  The labels beside each CCF refer to the visual phase computed from the
  GCVS period and the AAVSO ephemeris. The observing night number is 
  given on top of each panel according to Table~\ref{Tab:obs_log}. The 
  dotted line indicates the center-of-mass velocity
  (see Table~\ref{Tab:com_vr})}
  \label{Fig:scep_survey_K0_2}
\end{figure*}

Figure~\ref{Fig:scep_survey_K0_2} presents the sequence of CCFs obtained 
for S\,Cep. Several important features must be noted: 
(i) as seen on Fig.~\ref{Fig:scep_survey_K0_1}, the velocities of the red and
blue components are almost identical
from one cycle to the next (although their relative intensities change 
somewhat: compare the profiles at phases 0.95 and 1.96 on 
Fig.~\ref{Fig:scep_survey_K0_2});
(ii) the center-of-mass velocity derived from the submm CO lines falls in
between the red and blue peaks, as expected (Figs.~\ref{Fig:scep_survey_K0_1}
and \ref{Fig:scep_survey_K0_2}; see also 
Sect.\ref{Sect:com_vel});
(iii) the velocities derived from the near infrared CO lines (Hinkle et al.
1997) are very 
similar to those derived from the optical CCFs.

\subsection{Line doubling and center-of-mass velocity}
\label{Sect:com_vel}

The location of the center-of-mass (COM) velocity with respect to the
blue and red peaks provides an important information about the shock.
In the {\it classical} \sch\ scenario, the COM velocity falls in between 
the blue and red peaks. It might happen, however, that the shock is not 
strong enough to oppose the infalling matter. In that case, {\it both} 
peaks would be redshifted with respect to the COM velocity (indicating that 
the shock is receding in an Eulerian rest frame), as observed for RR\,Lyr
stars at some phases (Chadid \& Gillet 1996).

This does not occur in S\,Cep, as seen on Figs.~\ref{Fig:scep_survey_K0_1}
and \ref{Fig:scep_survey_K0_2} (see also Hinkle \& Barnbaum 1996) where
the dotted line corresponds to the COM velocity derived by Neri et al.\
(1998) from submm observations. And seemingly, {\it receding shocks are not 
observed in Miras at all}. Table~\ref{Tab:com_vr} lists the COM velocity 
(when available) for LPVs which exhibit a double-peak CCF. The first column 
gives the star name, the second column the velocities of the blue and red 
components (extreme value observed), the third column the median of the blue-
and red-peak velocities, the fourth column the COM velocity 
derived from submm observations and the last column the reference for this 
COM velocity.
The COM velocity does always fall in between the blue and red peaks (the
median of the red- and blue-peak velocities is in fact blueshifted with respect
to the COM velocity, as expected if the shock front is moving outwards), 
in agreement with the conclusion already reached by Hinkle et al.\ (1997)
from infrared CO lines. It is thus possible to conclude that, unlike in other 
classes of variable stars such as RR\,Lyrae, shock waves are rather strong 
in Mira stars.

\addtocounter{table}{+1}
\begin{table}
\caption[]{Center-of-mass velocity for stars exhibiting a 
double-peak CCF}
\begin{flushleft}
\begin{tabular}{rcccc}
\hline\noalign{\smallskip}
GCVS & CCF  &  median & Center-of-mass   & Ref.
\\ 
     & radial velocities & red/blue & velocity\\
     & (\kms)                & (\kms)          & (\kms)                   & \\
\noalign{\smallskip}
\hline\noalign{\smallskip}
 R And & $-$38/$-$10   & $-$24   & $-$20.3  & 2 \\
 R Cet & +22/+48       & +35     & +42.1    & 1 \\
 U Ari & $-$65/$-$43   & $-54$   & $-$46.5  & 1 \\
 R Gem & $-$57/$-$39   & $-$48   & $-$47.6  & 2 \\
 S CMi & +50/+72       & +61     & +65.9    & 2 \\
 W Aql & $-$54/$-$27   & $-$40.5 & $-$39.2  & 2 \\
 Z Cyg & $-$179/$-$160 & $-$169.5& $-$165.8 & 1 \\
AX Cep & $-$10/+8      & $-$1    & +0.4     & 1 \\
 S Cep & $-$39/$-$18   & $-$28.5 & $-$26.8  & 3 \\
\noalign{\smallskip}
\hline
\end{tabular}
\end{flushleft}
References:\\
1: Groenewegen et al.\ 1999 \\
2: Knapp et al.\ 1998 \\
3: Neri et al.\ 1998 \\
\label{Tab:com_vr}
\end{table}

\subsection{The key stellar parameter governing the occurrence of the 
\sch\ scenario}
\label{Sect:key_parameter}

\subsubsection{Statistics of the line-doubling phenomenon}
\label{Sect:statistics}

\begin{figure}
  \resizebox{\hsize}{!}{\includegraphics{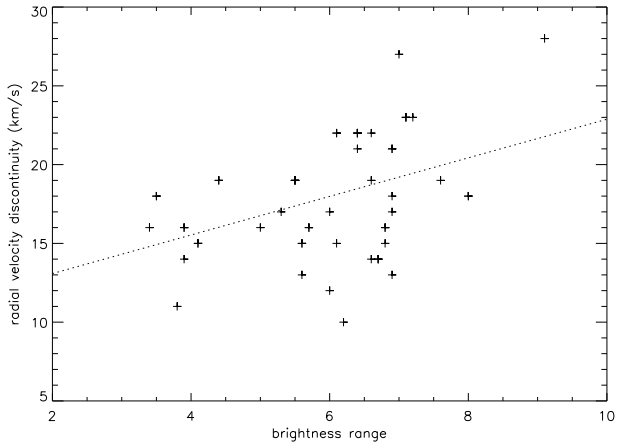}}
  \caption[]{Radial velocity discontinuity ($V_{\rm red} - V_{\rm blue}$)
  against the brightness range (expressed in mag) 
  for stars with a double 
  or asymmetrical peak when observed with the K0-template in the phase 
  range $-0.1$ -- 0.1. The 
  velocity discontinuity displayed corresponds to the maximum value
  recorded for the above phase range. The dashed 
  line is the regression line}
  \label{Fig:h_2p_dV_K0}
\end{figure}

\begin{figure}
  \resizebox{\hsize}{!}{\includegraphics{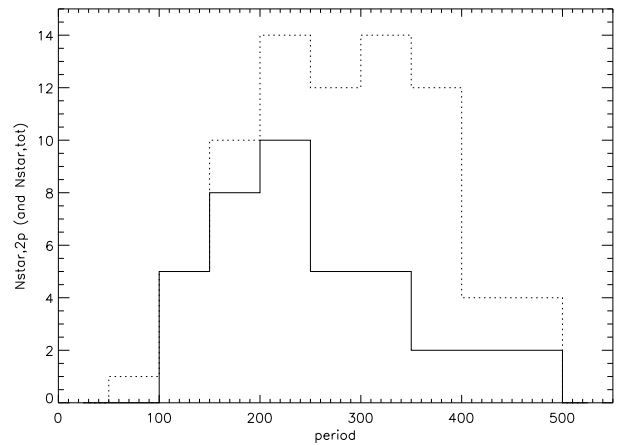}}
  \caption[]{Period distribution for stars with a double or asymmetrical 
  peak obtained with the K0-template (solid line) as compared to the 
  sample of stars observed at least once around maximum light (dashed line)}
  \label{Fig:h_2p_per_K0}
\end{figure}

\begin{figure}
  \resizebox{\hsize}{!}{\includegraphics{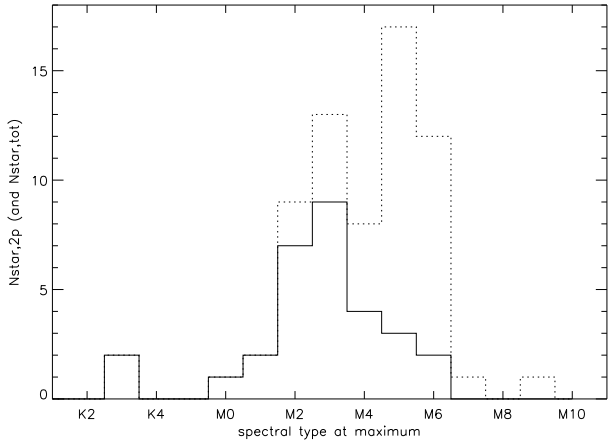}}
  \caption[]{Distribution of spectral types (at maximum light) for 
  oxygen-rich stars with a double or asymmetrical peak obtained with the 
  K0-template (solid line) as compared to the sample of stars observed at least
once around maximum light (dashed line)}
  \label{Fig:h_2p_spt_K0}
\end{figure}

Not all LPVs exhibit the line doubling (at least when considering only 
those CCFs classified as `2p' or `ap' as described in
Sect.~\ref{Sect:CCF}, thus excluding CCFs of the `2p?' and `np'
kinds; see Fig.~\ref{Fig:types_ccf}), 
even when observed at maximum light 
(see the example of X\,Oph in \p1): about 47\% of the sample stars observed 
at maximum light (between phases $-$0.10 and 0.10) did not exhibit line 
doubling with the K0-template (see Fig.~\ref{Fig:h_2p_phi_K0}). 
In an attempt to identify the key stellar parameter(s) governing the 
appearance of the line doubling, the distribution of several potentially
relevant parameters is displayed in Fig.~\ref{Fig:h_2p_dV_K0} 
($V_{\rm red} - V_{\rm blue}$ against the brightness range), 
Fig.~\ref{Fig:h_2p_per_K0} (period distribution) and 
Fig.~\ref{Fig:h_2p_spt_K0} (distribution of spectral type at maximum 
light for oxygen-rich LPVs).

There is a weak correlation between the brightness range
and the velocity discontinuity (Fig.~\ref{Fig:h_2p_dV_K0}), but 
the scatter in the relation is very large, albeit similar 
to that obtained by Hinkle et al. (1997; their Fig.~11) when
considering only Miras.
{\it
The important result from Fig.~\ref{Fig:h_2p_dV_K0} is that 
there is no visual-amplitude threshold beyond which line doubling would
only occur: instead, small-amplitude as well as large-amplitude Miras
are liable to exhibit the line-doubling phenomenon.}

The same conclusion is reached when considering the period
(Fig.~\ref{Fig:h_2p_per_K0}) or spectral type (Fig.~\ref{Fig:h_2p_spt_K0})
distributions: double-peak profiles are found among short- as well as 
long-period LPVs and among early- as well as late-type LPVs. The 
frequency of double-peak profiles is nevertheless somewhat larger among
short-period and early-type
LPVs (almost all LPVs with period less than 250~d and almost all 
oxygen-rich LPVs with a spectral type earlier than M3 showed line-doubling),
suggesting that the line-doubling is restricted to LPVs of small
radius (Sect.~\ref{Sect:radius}).

\subsubsection{The stellar radius: the key parameter?}
\label{Sect:radius}

Since the period is correlated with the luminosity and the spectral type 
with the effective temperature, and since these two parameters suffice
to define a stellar radius, we investigate the relationship between
the line-doubling phenomenon and the quantity
$\rho / R_{\sun} = [(L / L_{\sun}) (T/T_{\sun})^{-4} ]^{1/2}$.
Although this quantity is certainly related to the stellar radius, it
should not be taken at face value since the definition of a stellar radius
for LPVs is a complex task 
(e.g.\ Perrin et al.\ 1999; Hofmann et al.\ 2000). 
The quantity $\rho$ should somehow reflect the stellar size, at
least in a relative sense when comparing one star to another.

Luminosities were estimated from the 
$M_{\rm bol}$--$P$ relation of Feast et al.\ (1989), with no zero-point 
correction.

Effective temperatures of LPVs are also difficult to derive. Apart from a
scarce number of direct determinations from interferometric observations, little
is known about the temperature scale of Mira and semi-regular variables. 
Colour--temperature relations may in principle be used (e.g.\ Bessell et 
al.\ 1998; Alvarez et al.\ 2000b; Houdashelt et al.\ 2000), but in practice 
it is difficult to collect the required set of simultaneous photometric 
observations for a large sample of LPVs.

In order to estimate \Teff\ for a number of sample stars as large as possible, 
the choice was made to rely on a spectral type--temperature relation, 
taking advantage of the fact that almost all LPVs have been given a spectral 
type, at least at maximum light. The spectral type/temperature scale of 
Perrin et al.\ (1998) was used for this purpose, with the spectral types at 
maximum light taken from the GCVS.
As spectral-type assignments generally carry some degree of subjectivity, so 
will the \Teff\ values derived from the present procedure. 
Their accuracy will be sufficient for the present purpose of identifying a
possible relationship between $\rho$ and the line-doubling phenomenon.
The derived $\rho$ values are listed in Table~3.

\begin{figure}
  \resizebox{\hsize}{!}{\includegraphics{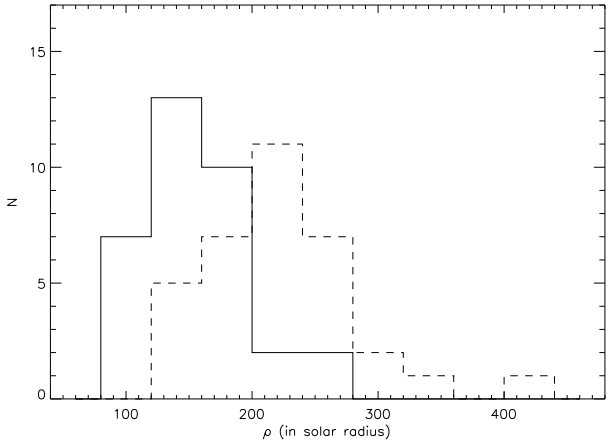}}
  \caption[]{Distribution of the stellar size $\rho$ (see text) for
  stars with a double or asymmetrical peak obtained with the K0-template 
  (full line) and for stars for which a double or asymmetrical peak has 
  not been obtained even when observed at maximum light (dashed line)
}
  \label{Fig:h_radius}
\end{figure}

Figure~\ref{Fig:h_radius} shows the $\rho$ distribution for two sub-samples: 
(i) the stars for which a double or asymmetrical peak was obtained 
with the K0-template (solid line);
(ii) the stars for which a double or asymmetrical peak was not  
obtained, even when observed at maximum light (dashed line). 
Semi-regular variables (as they do not follow the period-luminosity 
relation) and carbon stars (as they do not obey the oxygen-rich spectral 
type/temperature relation) are excluded from Fig.~\ref{Fig:h_radius}. 
Stars with a double peak are clearly separated from single-peak stars.
A large majority of small-size stars ($\rho < 200$~R$_{\sun}$) present a double
peak, whereas large-size stars ($\rho > 200$~R$_{\sun}$) are almost all 
single-peak stars. Such a segregation is quite remarkable given the various
sources of uncertainty on $\rho$. It accounts for the results of \p1,
namely the occurrence of line-doubling in RT\,Cyg ($\rho = 130$~R$_{\sun}$) 
but not in X\,Oph ($\rho = 216$~R$_{\sun}$).

 Fig.~\ref{Fig:h_radius}, which shows that the line-doubling phenomenon
is only present in the most compact Mira stars, is one of the
most striking result of this paper.
At this stage, two interpretations of this result are still possible:\\
(i) it is an artefact of the use of the K0III template which is not
able to reveal the double peaks  in the
LPVs with large $\rho$ which have very late and crowded spectra;\\
(ii) the observed segregation in terms of $\rho$ is real and related
to the physics of the shocks in LPVs.
\medskip\\
These two possibilities are discussed in turn in the remaining of this 
section.
\medskip\\
{\it \thesubsubsection.1 An artefact of the method?}
\medskip\\
\noindent 
Since the K0III
template better matches the spectra of early-type LPVs than of
late-type LPVs, 
it may possibly be more efficient
in detecting line-doubling in early-type LPVs, thus possibly causing a bias
against the detection of line doubling in large-$\rho$ LPVs.
The interpretation of Fig.~\ref{Fig:h_radius} simply in terms of such
a selection bias caused by the use of too warm a template appears
however too simplistic, as we now discuss. 

\begin{figure}
  \resizebox{\hsize}{!}{\includegraphics{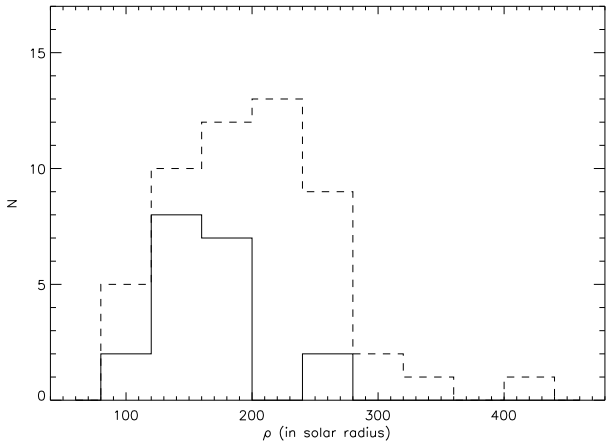}}
  \caption[]{
  Same as Fig.~\protect\ref{Fig:h_radius} for the M4-template
}
  \label{Fig:h_radius_M4}
\end{figure}

\begin{figure*}
  \resizebox{\hsize}{!}{\rotatebox{90}{\includegraphics{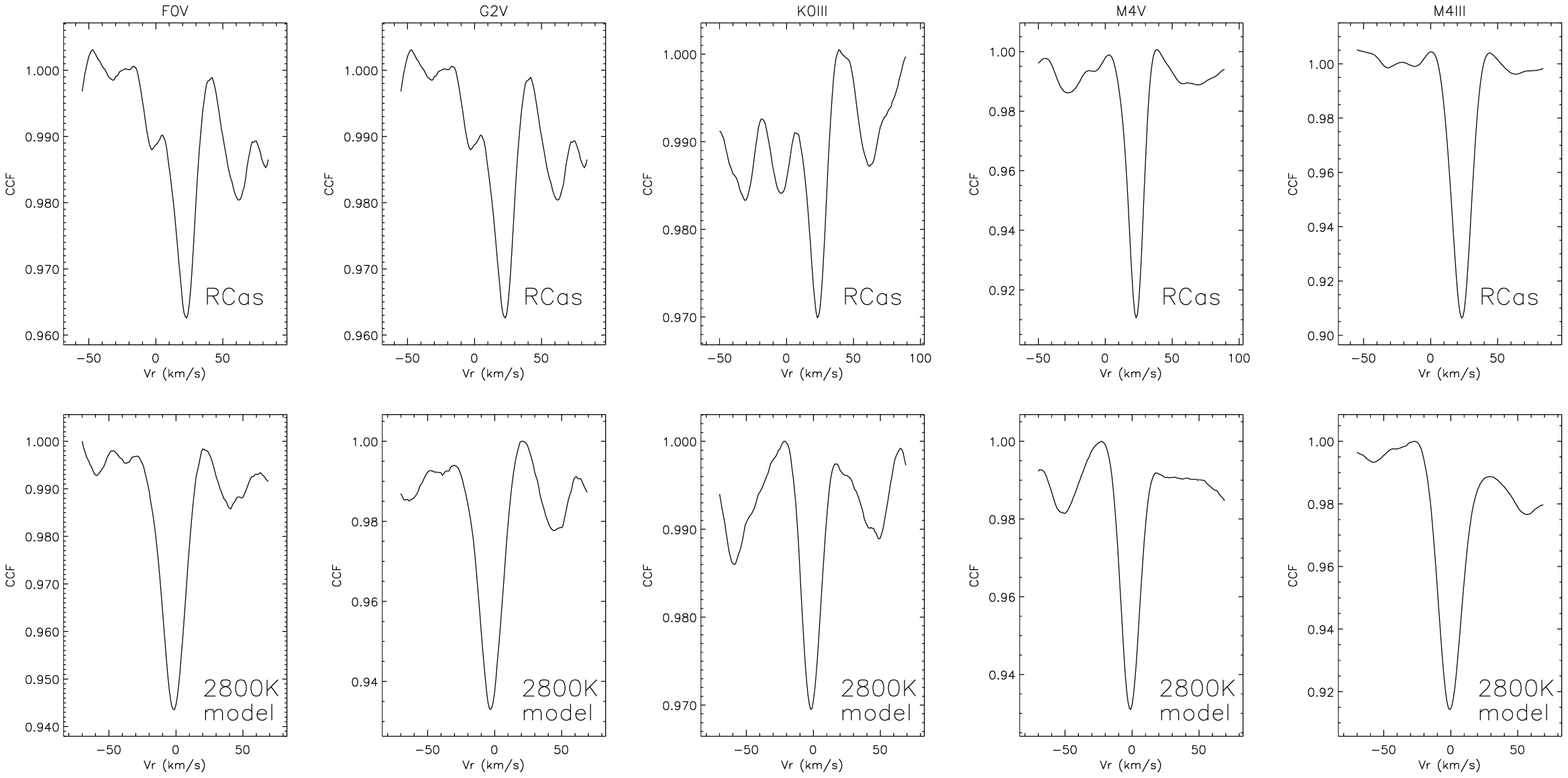}}}
  \caption[]{
Upper row: CCFs of R~Cas (N13, phase 0.02), a late-type LPV (M6e-M10e), 
obtained with various templates (F0V, solar G2V, K0III, M4V
and M4III).\\
Lower row: CCFs obtained with the same templates applied on a 2800~K
synthetic spectrum (see Sect.~3.1 of Paper~II). 
Note how the CCFs of R Cas are
strikingly similar (especially their far wings) to those of the 
synthetic spectrum at 2800~K, with the exception
of the secondary blue peak observed at about 0~\kms\ in the F0V,
G2V and K0III CCFs of R~Cas. This peak might thus tentatively 
be attributed to the dynamics of the envelope of R~Cas, since a
corresponding peak is not present in the CCF of a static model atmosphere. 
}
  \label{Fig:RPsc}
\end{figure*}

Since the cooler M4V template\footnote{We show below -- see
  Fig.~\protect\ref{Fig:RPsc} -- that the use of a M4III template does 
  not improve the situation, since it provides CCFs that are very
  similar to those obtained with the M4V template} 
constitutes a better match to the crowded
spectra of late-type LPVs, one might have
expected a better efficiency to detect line doubling with that template. 
We already
showed in Sect.~\ref{Sect:CCF} that even {\it less}
double-peaked
CCFs are found with the M4V template than with the
K0III one. Fig.~\ref{Fig:h_radius_M4} shows that the segregation of
single/double peaks in terms of $\rho$ remains basically the same with 
the M4V template as with the K0III template.

Nevertheless, by considering infrared CO $\Delta v = 3$ lines around
1.6 $\mu$m, Hinkle et al. (1984, 1997) reported line doubling around
maximum light for several 
{\it  late-type }
LPVs
with large $\rho$ 
(Table~\ref{Tab:Hinkle}) that were flagged as 1p or 1p? with the K0III 
and M4V templates.  
{\it  Thus line doubling is definitely present at least in 
some of those stars as well.
}

Quite interestingly, Hinkle et al. (1984)
note that in these late-type LPVs line doubling is more easily seen
in the deep photospheric layers probed by the  
infrared CO $\Delta v = 3$ lines around
1.6 $\mu$m than in the higher layers where the blue-violet absorption
lines form. In an attempt to uncover double-peak CCFs  in late-type
LPVs as well, their spectra were correlated with various masks 
(constructed with the procedure of Baranne et al. 1979)
mimicking F0V, G2V (solar), K4III (81 Gem) and M4III
(HD~123657) spectra.  

While the cool  masks (M4V and M4III) 
yield very clean, albeit single, CCFs, interesting
results are indeed obtained with the warm F0V and G2V masks, which
are expected to probe deeper layers. They 
reveal asymmetric, or even double-peak, CCFs (Table~\ref{Tab:Hinkle} and
Fig.~\ref{Fig:RPsc}), with velocities similar to those obtained by
Hinkle et al. (1984). There is however a drawback to the
application of warm templates to cool LPVs. Since the warm templates
contain far fewer lines than the spectrum of cool LPVs, many spurious
secondary peaks must be expected in their CCFs. These peaks correspond 
to the chance coincidence occurring between the mask and the forest of 
lines flanking the lines actually probed by the mask.  This situation
is well illustrated by Fig.~\ref{Fig:RPsc} which compares the CCFs
obtained with different masks applied on the late-type LPV R Cas and
on the synthetic spectrum of a 
{\it  static }
2800~K model. The
latter CCFs make it possible to identify the spurious secondary peaks
caused by the correlation noise invoked above. By comparison, the CCFs
of 
R~Cas obtained with the warm templates exhibit the same spurious peaks 
as the 2800~K model, except that a supplementary blue component
appears around 0~\kms\ (see also Table~\ref{Tab:Hinkle}).  
This component not observed in the CCF of a static spectrum 
may thus tentatively be related to the dynamics of the
envelope of R Cas. 

\begin{table*}
\caption{\label{Tab:Hinkle}
Late-type LPVs for which line doubling has been observed by Hinkle et
al. (1984) using CO $\Delta v = 3$ lines around 1.6
$\mu$m: Comparison of the infrared velocities with those derived from
various CCFs at about the same phase
}
\begin{tabular}{lll||lc||lc|lc|lc|lc}
\hline
   &             &&  \multicolumn{2}{c}{CO $\Delta v = 3$}
                 & \multicolumn{2}{c}{CCF} 
                 & \multicolumn{2}{c}{K0III }
                 & \multicolumn{2}{c}{M4V }
                 & \multicolumn{2}{c}{F0V }
\\
\hline
\footnotesize
   & Spect. type & $\rho$ (R$_\odot$)&  Phase & $V_r$ 
                 & Phase & night & & $V_r$ & & $V_r$ & & $V_r$ \\
\hline   
R And & S3,5e - S8,8e & 217 & 0.96 & -29.3/-5.7 & 0.98 & N26 & 2p & -38.3/-10.2 & 1p & -9.5 & 2p & -38.1/-9.8\\
R Cas & M6e - M10e & 282 & 0.03 & +5.9/+28.5  & 1.02 & N13 & 1p? &- & 1p & +22.7& 2p? & 0/+21.7 \\
T Cep & M5.5e - M8.8e & 238 & 0.87 & -19.2/-2.4 & 1.91 & N23 & np &- & 1p & -14.9 & 1pa & -19.5/-10.3 \\ 
o Cet & M5e - M9e & 150 &  0.96 & +44.7/+67.8 & 1.91 & N15 & 1p? & - & 1p & +61.7 & 2p? & +38.4/+57.9 \\
\hline

\end{tabular}
\medskip\\
\end{table*}

Many of the LPVs with large $\rho$ that were originally classified as
1p or 1p? from their K0III CCF actually behave like R Cas as shown on
Fig.~\ref{Fig:RPsc}. 

  In conclusion, double-peaked
  CCFs may actually be present in LPVs with large $\rho$ as well.
We nevertheless stress that this conclusion must be subject to
caution, since these double-peak CCFs -- if real --
are quite difficult to detect because of the crowded
  nature of their parent spectra. 
In Sect.~\ref{Sect:radius}.2, we discuss the possibility that this
difficult detection may also be caused partly by the 
secondary peaks being truly weaker in stars
with large $\rho$ due to weaker shocks.
\medskip\\
\noindent{\it \thesubsubsection.2 Weaker shocks in stars with large $\rho$?}
\medskip\\
\noindent 
The behaviour of the H$\delta$ emission line in our sample of
  LPVs provides important clues to interpret in terms of the properties
  of the shocks in LPVs, the result that the 
line-doubling phenomenon
is conspicuous only in the most compact LPVs.

\begin{figure}
  \resizebox{\hsize}{!}{\rotatebox{90}{\includegraphics{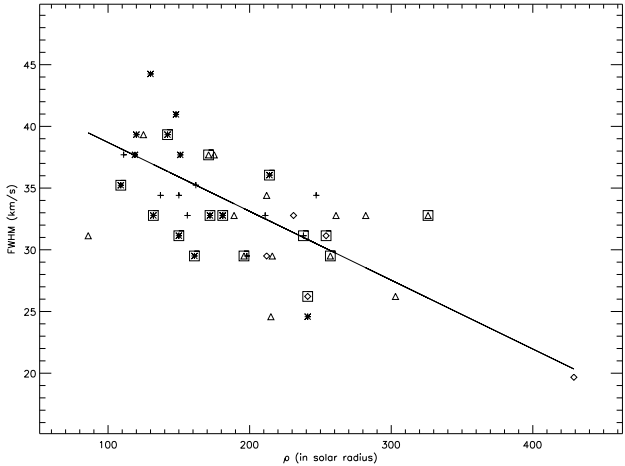}}}
  \caption[]{
Full width at half maximum of the H$\delta$ emission line as a
function of the stellar size $\rho$. The solid line is a least-square
fit through the data. The different symbols refer to
the different types of CCFs obtained with the K0III template as defined in
Table~\protect\ref{Tab:ccf}: 
asterisks = 2p, crosses = ap, triangles = 1p, diamonds = np. A large
square around the previous symbols denotes stars with a double
H$\delta$ profile. 
The FWHM for these profiles has
been measured under the assumption that 
the self-reversal model applies to them. This model 
supposes that the emission profile is mutilated by a true
absorption line caused by hydrogen located above the shock
in the radiative precursor (Gillet 1988b). 
The `true' FWHM is that corresponding to the non-mutilated emission line. 
It has been estimated along the guidelines (involving gaussian fits) 
described in Fig.~5 of Gillet (1988b). The accuracy of such measurements is 
estimated to be about 10\%
}
\label{Fig:Hd}
\end{figure}

The first important result regarding H$\delta$
is that it appears in emission around maximum light 
for all LPVs in our sample 
(see Fig.~\ref{Fig:Hd} discussed
below). Since H$\delta$ is considered as a good signature of the
presence of shocks in those atmospheres,  
the above result thus suggests that double-peak CCFs, being another signature 
of shocks, should be observed as well in {\it all} 
LPVs.
We argue in the present section that the behaviour of H$\delta$
and the statistics of double-peak CCFs in LPVs may be reconciled 
by realizing that
the shock strength
becomes smaller in stars of larger $\rho$.
Fig.~\ref{Fig:Hd} presents the full width at
half maximum (FWHM) of the H$\delta$ emission line vs. the stellar size 
$\rho$. We observe a trend of decreasing H$\delta$ FWHMs 
with increasing stellar sizes $\rho$. 
Among the Balmer lines, H$\delta$ is the best 
indicator of the energy transferred by the shock to the gas,
since H$\delta$ suffers the least from 
mutilations induced by molecular absorptions 
(the TiO line absorption and H$_{2}$ scattering
coefficients are indeed minimum around H$\delta$; Gillet 1988b).
Its width measures the temperature in the post-shock gas 
where the Balmer
emission lines form. The heating in the shock wake must in turn be
related to the energy carried by the shock.  

Similarly, the difficulty of detecting double-peaked CCFs 
in stars with large $\rho$  might also somehow relate to
the weakness of the shocks in those stars (as inferred from their 
smaller H$\delta$ FWHMs), although a definite proof
of this statement would require to solve the equation of radiative
transfer in dynamical atmospheres, with detailed shock models.

Self-consistent dynamical models of LPV atmospheres are clearly 
necessary to
confirm that 
the strongest shocks only occur in the more compact Mira stars.
In any case, the relation of the shock strength 
with the stellar size empirically revealed by
Fig.~\ref{Fig:Hd} will have to be accounted
for by future dynamical models of LPV atmospheres.

Furthermore, the predictive power of Fig.~\ref{Fig:h_radius} is of interest
in relation with abundance analyses of LPVs in external systems, where
the available spectral resolution may not be high enough to reveal the
line doubling. This may introduce large errors on the abundances, and
Fig.~\ref{Fig:h_radius} reveals that such a risk is highest for warm and
short-period LPVs.
  
\subsection{Repeatability of the line doubling from one cycle to the other
and search for binaries}
\label{Sect:binary}

\begin{figure}
  \resizebox{\hsize}{!}{\includegraphics{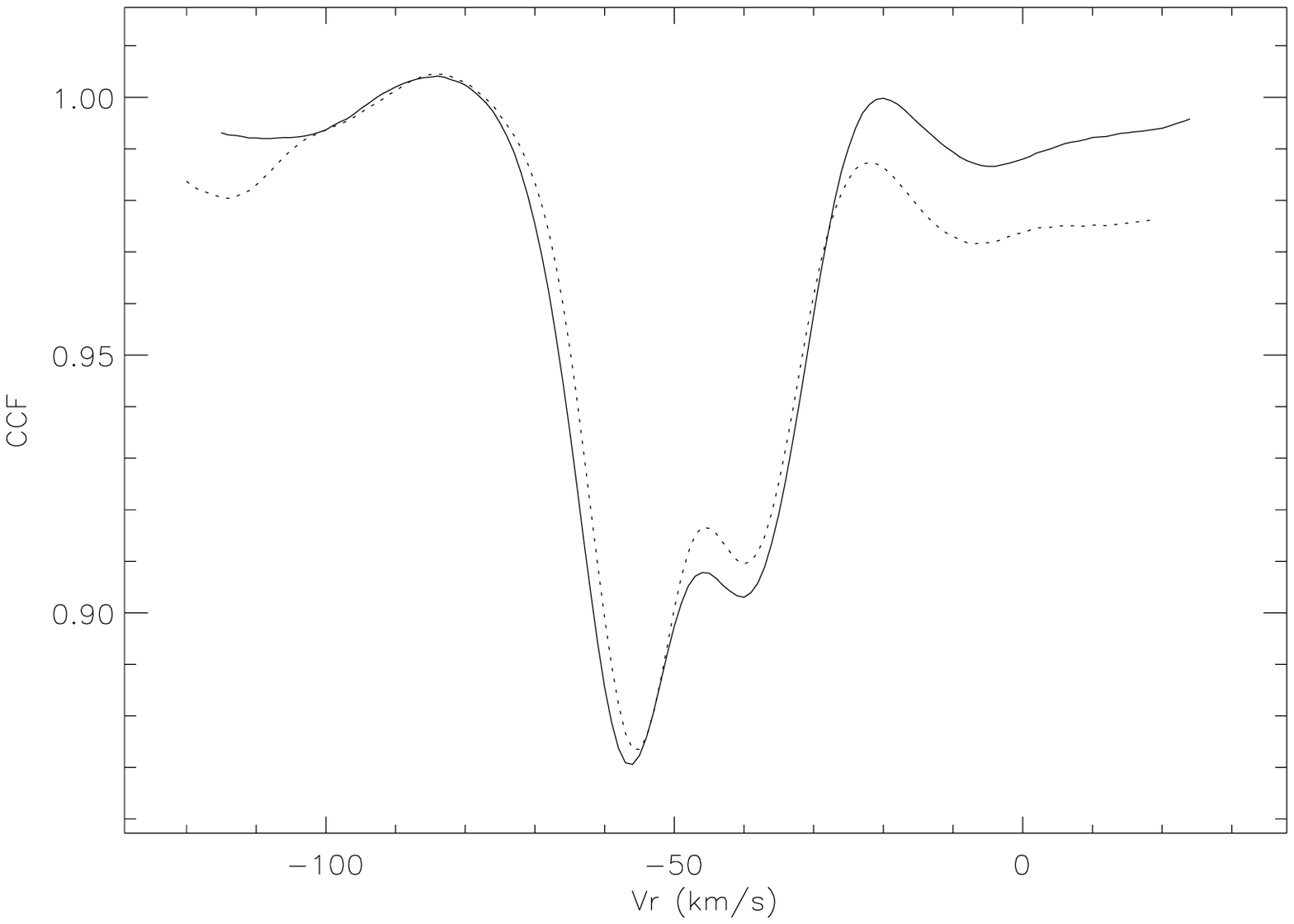}}
  \caption[]{Cross-correlation profiles of R\,Gem obtained with the default 
  K0\,III mask at phases 0.99 (solid line) and 2.09 (dotted line).
  Velocities are ($V_{\rm blue}$, $V_{\rm red}$) =  
  ($-$56.6, $-$38.8)~\kms\ (phase 0.99) and ($-$55.8,$-$39.2)~\kms\ 
(phase 2.09)}
  \label{Fig:rgem_2phi}
\end{figure}

\begin{figure}
  \resizebox{\hsize}{!}{\includegraphics{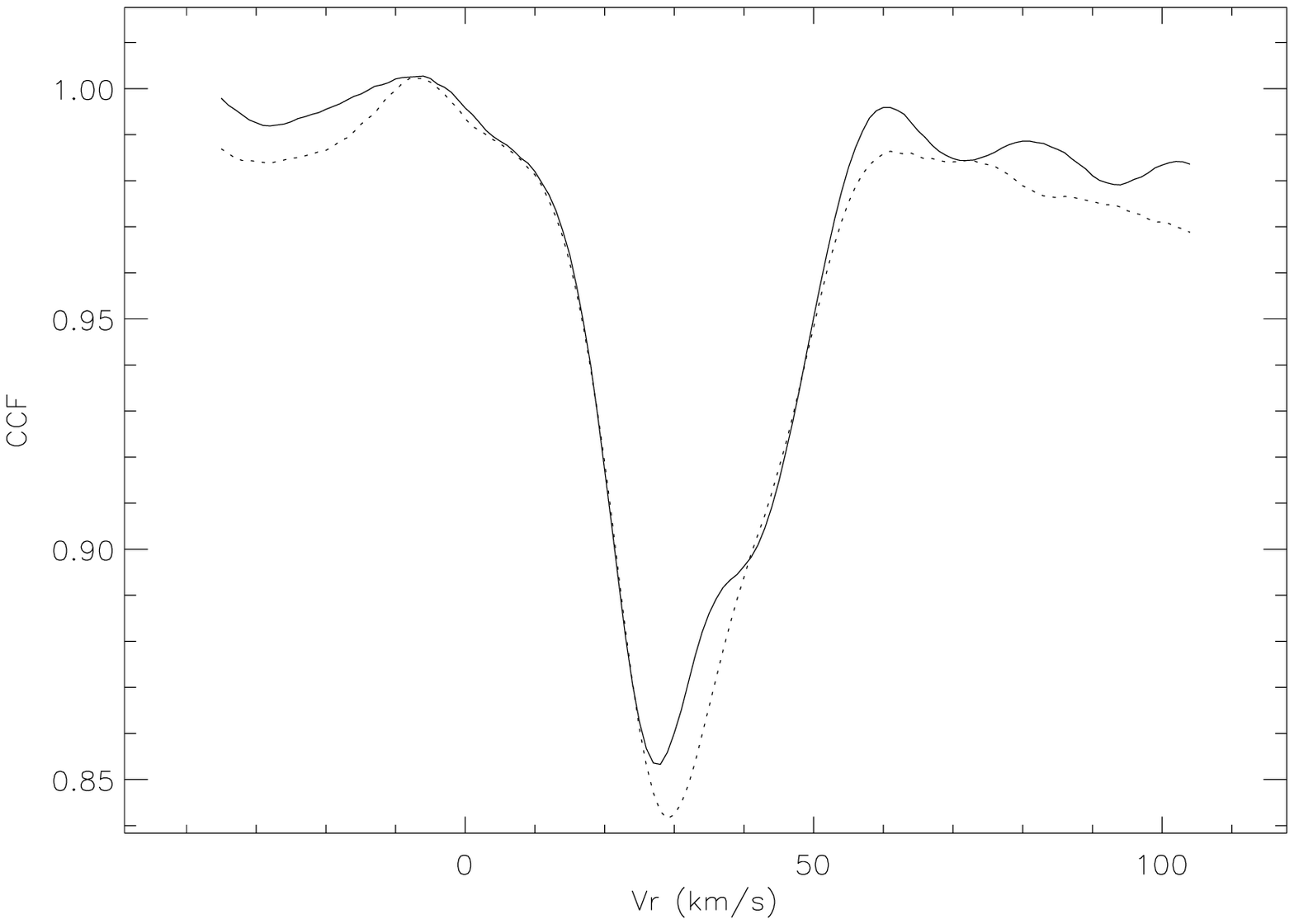}}
  \caption[]{Cross-correlation profiles of ST\,And obtained with the default 
  K0\,III mask at phases 0.99 (solid line) and 1.98 (dotted line).
  Velocities are ($V_{\rm blue}$, $V_{\rm red}$) = 
  (+27.5, +43.1)~\kms\ (phase 0.99) and (+28.7, +43.5)~\kms\ (phase 1.98)}
  \label{Fig:stand_2phi}
\end{figure}

Observations spread over several cycles are available for some stars
and allow to address the questions of (i) the repeatability of the line
doubling from one cycle to the other (as far as the relative peak 
intensities and velocities are concerned), and (ii) the frequency of 
spectroscopic binaries among LPVs. In Figs.~\ref{Fig:rgem_2phi} and 
\ref{Fig:stand_2phi} are presented the CCFs obtained for R\,Gem and ST\,And 
respectively, at maximum light for two consecutive pulsational cycles. 
The CCF profile is remarkably similar from one cycle to the other, with the
same velocities to within 1~\kms\ for the blue {\it and} red peaks.
This result, based on only two stars, may not necessarily be
extrapolated to all LPVs.

\begin{figure}
  \resizebox{\hsize}{!}{\includegraphics{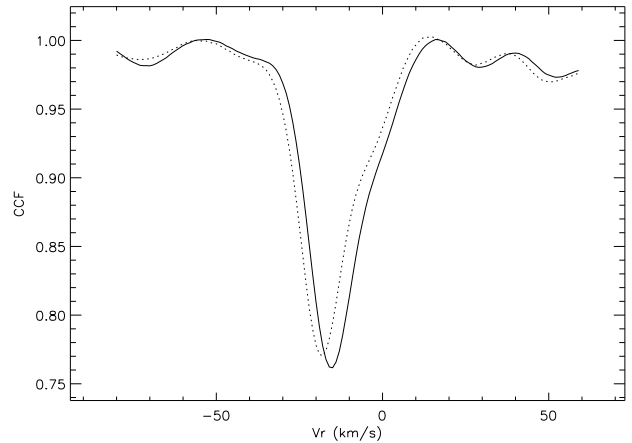}}
  \caption[]{Cross-correlation profiles of S\,Cam obtained with the default 
  K0\,III mask at phases 0.15 (solid line) and 1.13 (dotted line).
  Velocities are ($V_{\rm blue}$, $V_{\rm red}$) =  
  ($-$15.4, $-$0.9)~\kms\ (phase 0.15) and ($-$18.2, $-$3.1)~\kms\ (phase 1.13)}
  \label{Fig:scam_2phi}
\end{figure}

\begin{figure}
  \resizebox{\hsize}{!}{\includegraphics{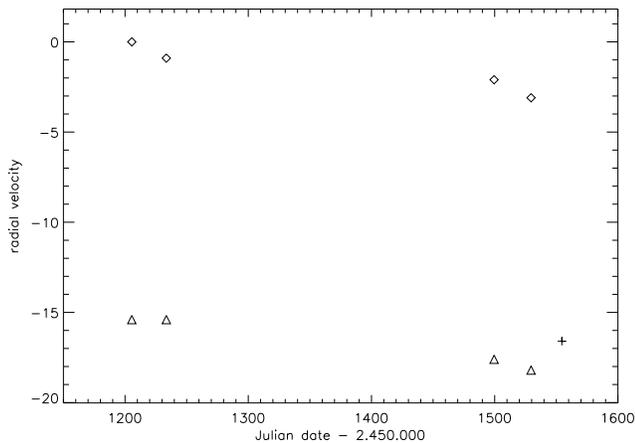}}
  \caption[]{Radial-velocity curve of S\,Cam with the default K0\,III 
  mask. Symbols are as in the last panel of Fig.~\ref{Fig:scep_survey_K0_1}.
  Notice the long-term trend, unlike the situation observed for S\,Cep in
  Fig.~\ref{Fig:scep_survey_K0_1}}
  \label{Fig:scam_vr}
\end{figure}

By contrast, the situation observed in S\,Cam (Fig.~\ref{Fig:scam_2phi}) is 
quite interesting: although the CCF shape is almost identical for similar
phases of two consecutive cycles, it is shifted by about 3~\kms.
Fig.~\ref{Fig:scam_vr} shows the radial-velocity curve of S\,Cam: the common
velocity shift for the blue and red peaks from one cycle to the next is quite
clear, and flags this star as a suspected binary. The example
 of S~Cam  only serves to illustrate the method, since to confirm the
 binary nature of S~Cam would still require  
 to prove that the phases compared on Fig.~\ref{Fig:scam_2phi} are really
 identical, i.e. that the period is constant (this is not
 necessarily the case for a SRa variable like S Cam) and
 that the observed radial-velocity shift is larger than 
 the typical cycle-to-cycle scatter.

Little is known about the binary frequency among LPVs (see Jorissen 2001 for a
recent review). As pointed out in 
Sect.~\ref{Sect:temporal_evol} and Fig.~\ref{Fig:scep_survey_K0_1}, the
radial-velocity variations induced by the LPV pulsation are large 
($\Delta V_{\rm r} \sim$ 20~\kms\ over $\sim$ 1 y) as compared to the 
variations expected for the orbital motion of a large giant in a relatively 
wide binary system ($\Delta V_{\rm r} \sim$ 5~\kms\ over several years). 

However, the example of S\,Cam provides a hint on how binary systems may
be found among LPVs exhibiting a double-peak CCF: a similar
velocity shift observed for both the red and blue components from one cycle
to the other may be indicative of a binary motion.
Such a strategy is unfortunately quite time-consuming as only double-peak
CCFs obtained around the same pulsational phase in successive cycles may
be used, and is obviously not applicable to those LPVs exhibiting only a
single-peak CCF.

In our sample of 81 LPVs, 43 showed at least once a double-peak CCF with 
the K0-template (Table~3), but only 9 could be observed at about the same phase
(within 0.1) in consecutive cycles and exhibited double-peak CCFs (namely
Y\,And, V\,Tau, S\,Cam, R\,Gem, Y\,Vir, Z\,Oph, RT\,Cyg, S\,Cep and 
ST\,And).
In this sample, the average absolute velocity shift between similar phases in
consecutive cycles amounts to 
$\overline{\mid \Delta V_{\rm blue} \mid} = 1.3 \pm 0.6$ (r.m.s.) \kms\ 
and $\overline{\mid \Delta V_{\rm red} \mid} = 1.4 \pm 1.1$ (r.m.s.) \kms\
when removing the suspected binaries S\,Cam 
(having $\mid \Delta V_{\rm blue} \mid = 2.2$ and 2.8~\kms, 
and $\mid \Delta V_{\rm red} \mid = 2.1$ and 2.2~\kms) and, somewhat
less likely, Y~Vir (having $\mid \Delta V_{\rm blue} \mid = 3.0$~\kms, 
and $\mid \Delta V_{\rm red} \mid = 2.2$~\kms) .
No other star in the above list shows a significant signature of binary motion.

For the stars not suspected of being binaries, the red- and blue-peak 
shifts from one cycle to the next may sometimes be quite different, since the
average 
$\mid \Delta V_{\rm blue} - \Delta V_{\rm red} \mid$, instead of being close
to 0, amounts to $1.8 \pm 1.5$ (r.m.s.) \kms.

\section{Summary}
\label{Sect:conclusion}

The most striking result relative to the 
statistics of occurrence of the line-doubling phenomenon in a sample of
81 LPVs of various periods, spectral types and
brightness ranges is the fact that the compact LPVs are more 
prone to exhibit the line-doubling phenomenon than large-size LPVs
(at 
least when comparing their K0III CCFs). 
The possibility that this segregation be an artefact of the use of the K0III
template may not be totally excluded, since
warmer masks (F0V and G2V) applied to the most extended and 
  coolest LPVs yield asymmetric
  cross-correlation functions. This suggests that line doubling is
  occurring in those stars as well.  Although a firm conclusion on this
  point is hampered by the large correlation noise present in the
  CCFs of cool LPVs obtained with warm masks, the occurrence of line
  doubling in those stars is confirmed by the double
  CO $\Delta v = 3$ lines observed around 1.6 $\mu$m by Hinkle
  et al. (1984).
Moreover, the H$\delta$ line in emission, which is another signature of the
presence of shocks, is observed as well 
in the most extended stars, 
although with a somewhat narrower profile. This is an indication that the shock is
weaker in extended than in compact LPVs, which may also contribute to
the difficulty of detecting line doubling in cool, extended LPVs.

Stars with double absorption lines around maximun light exhibit S-shaped
radial-velocity curves similar to the one observed for infrared lines.

A comparison with the center-of-mass (COM) velocity obtained
from submm CO lines originating in the circumstellar envelope reveals that the
median velocity between the red and blue peaks is blueshifted with respect to
the COM velocity, as expected if the shock moves upwards. 

\begin{acknowledgements}
RA benefits of a TMR "Marie Curie" Fellowship at ULB. A.J.\ is Research 
Associate from the {\it Fonds National de la Recherche Scientifique} 
(Belgium). We thank J. Mattei for providing data from the AAVSO database. 
This program would not have been possible without the generous
allocation of telescope time at the {\it Observatoire de
  Haute-Provence} (operated by C.N.R.S., France). We tank S. Udry for
providing us with F0V and G2V templates.
\end{acknowledgements}


\addtocounter{table}{-3}
\renewcommand{\baselinestretch}{0.7}
\begin{table*}
\caption[]{Results: heliocentric radial velocities (in km~s$^{-1}$,
between parenthesis) and CCF types obtained with the K0- and
M4-templates, following the coding defined in the text
(1p: single peak; 2p: double peak; ap: asymmetrical peak;
1p?: doubtful single peak; 2p?: doubtful double peak;
 np: noisy profile). The column labelled $\rho$ provides an estimate
 of the stellar radius (see text) for oxygen-rich Mira stars only.}
\small{
\begin{flushleft}
\begin{tabular}{lllllll}
\hline\noalign{\smallskip}
Name & $\rho$      & Night  & Julian Date & Phase & 
K0-template & M4-template \\
     & ($R_\odot$) & Number & (2450000+)  &       & 
            &             \\
\noalign{\smallskip}
\hline\noalign{\smallskip}
        SV And & 211 & 
      N2 & 1060.5 &  0.04 &           1p ($-93.1$) &           1p ($-91.8$) \\
 & & N13 & 1386.5 &  1.08 &                    2p? &           1p ($-90.2$) \\
         T And & 181 & 
      N4 & 1093.5 &  0.95 &   2p ($-95.5$/$-79.2$) &           1p ($-83.0$) \\
 & & N15 & 1450.5 &  2.11 &                     np &           1p ($-84.8$) \\
         R And & 217 & 
     N26 & 1766.5 &  0.98 &   2p ($-38.3$/$-10.2$) &            1p ($-9.5$) \\
         Y Cep & 217 & 
     N26 & 1766.5 &  0.00 &                     np &            1p ($-6.0$) \\
         U Psc & 135 & 
     N27 & 1767.5 &  0.09 &           1p ($+37.5$) &           1p ($+36.8$) \\
         R Psc & 196 & 
      N4 & 1093.5 &  0.03 &           1p ($-45.4$) &           1p ($-46.0$) \\
 & & N15 & 1450.5 &  0.98 &           1p ($-44.5$) &           1p ($-45.0$) \\
 & & N16 & 1499.5 &  1.12 &                     np &           1p ($-45.5$) \\
 & & N17 & 1529.5 &  1.20 &           1p ($-46.8$) &           1p ($-45.4$) \\
 & & N27 & 1767.5 &  1.90 &           1p ($-43.6$) &           1p ($-44.4$) \\
         Y And & 150 & 
      N2 & 1060.5 &  0.03 &    2p ($-15.8$/$+0.9$) &            1p ($-4.0$) \\
 & &  N3 & 1085.5 &  0.14 &    ap ($-13.7$/$-1.2$) &            1p ($-4.5$) \\
 & & N16 & 1499.5 &  2.02 &    2p ($-16.5$/$+4.7$) &            1p ($+2.8$) \\
 & & N17 & 1529.5 &  2.15 &    2p ($-14.8$/$+3.1$) &            1p ($+2.1$) \\
 & & N18 & 1554.5 &  2.27 &                     np &            1p ($+2.2$) \\
       $o$ Cet & 150 & 
      N4 & 1093.5 &  0.92 &                    2p? &           1p ($+59.1$) \\
 & & N15 & 1450.5 &  1.91 &                    1p? &           1p ($+61.7$) \\
 & & N17 & 1529.5 &  2.14 &           1p ($+59.7$) &           1p ($+61.6$) \\
         R Cet & 132 & 
      N3 & 1085.5 &  0.07 &   2p ($+31.7$/$+48.2$) &           1p ($+45.2$) \\
 & &  N4 & 1093.5 &  0.30 &                    2p? &           1p ($+46.1$) \\
 & &  N7 & 1233.5 &  0.96 &           1p ($+45.5$) &           1p ($+44.1$) \\
 & & N13 & 1386.5 &  1.88 &           1p ($+47.5$) &           1p ($+47.4$) \\
 & & N15 & 1450.5 &  2.26 &   2p ($+21.7$/$+44.5$) &           1p ($+46.0$) \\
 & & N18 & 1554.5 &  2.89 &                    1p? &           1p ($+45.0$) \\
 & & N19 & 1597.5 &  3.15 &                    1p? &           1p ($+44.8$) \\
         U Cet & 148 & 
     N27 & 1767.5 &  0.01 &                     np &           1p ($-38.3$) \\
         R Tri & 175 & 
     N15 & 1450.5 &  0.30 &                     np &   ap ($+53.3$/$+65.3$) \\
 & & N17 & 1529.5 &  0.60 &                     np &           1p ($+58.6$) \\
 & & N19 & 1597.5 &  0.85 &           1p ($+59.7$) &           1p ($+61.0$) \\
         U Ari & 214 & 
      N2 & 1060.5 &  0.00 &                    2p? &           1p ($-43.8$) \\
 & &  N3 & 1085.5 &  0.07 &                     np &           1p ($-44.2$) \\
 & & N15 & 1450.5 &  1.05 &           1p ($-40.4$) &           1p ($-40.3$) \\
 & & N17 & 1529.5 &  1.26 &   2p ($-64.6$/$-42.9$) &           1p ($-40.3$) \\
 & & N18 & 1554.5 &  1.33 &                     np &           1p ($-40.0$) \\
        SS Cep &     & 
     N15 & 1450.5 & 1.56: &           1p ($-49.0$) &           1p ($-49.0$) \\
 & & N18 & 1554.5 & 2.71: &           1p ($-48.3$) &           1p ($-48.3$) \\
 & & N19 & 1597.5 & 3.19: &           1p ($-48.2$) &           1p ($-47.7$) \\
         R Tau & 212 & 
      N3 & 1085.5 &  0.01 &           1p ($+31.6$) &           1p ($+32.1$) \\
 & & N15 & 1450.5 &  1.15 &                     np &           1p ($+27.0$) \\
 & & N17 & 1529.5 &  1.40 &                     np &           1p ($+29.2$) \\
 & & N20 & 1598.5 &  1.61 &                     np &           1p ($+29.4$) \\
         V Tau & 111 & 
      N4 & 1093.5 &  0.08 &   2p ($+57.5$/$+76.8$) &   2p ($+57.2$/$+74.8$) \\
 & & N15 & 1450.5 &  2.01 &   2p ($+56.0$/$+77.8$) &           1p ($+73.8$) \\
 & & N17 & 1529.5 &  2.48 &                     np &           1p ($+74.0$) \\
 & & N20 & 1598.5 &  2.89 &           1p ($+77.4$) &           1p ($+77.4$) \\
         R Aur & 292 & 
      N4 & 1093.5 &  0.54 &                     np &            1p ($-2.5$) \\
 & & N15 & 1450.5 &  1.25 &                     np &            1p ($+5.8$) \\
 & & N20 & 1598.5 &  1.57 &                     np &            1p ($+4.4$) \\
         W Aur & 171 & 
      N3 & 1085.5 &  0.95 &          1p ($-131.3$) &          1p ($-132.6$) \\
 & & N19 & 1597.5 &  2.82 &                     np & ap ($-140.5$/$-130.4$) \\
        RU Aur & 326 & 
      N3 & 1085.5 &  0.97 &                     np &           1p ($-21.9$) \\
 & & N18 & 1554.5 &  1.98 &           1p ($-18.9$) &           1p ($-20.1$) \\
 & & N19 & 1597.5 &  2.07 &                     np &           1p ($-19.5$) \\
         S Cam &     & 
      N6 & 1205.5 &  0.07 &    2p ($-15.4$/$+0.0$) &                     np \\
 & &  N7 & 1233.5 &  0.15 &    2p ($-15.4$/$-0.9$) &                     np \\
 & & N17 & 1529.5 &  1.06 &    2p ($-17.6$/$-2.1$) &                     np \\
 & & N18 & 1554.5 &  1.13 &    2p ($-18.2$/$-3.1$) &                     np \\
 & & N19 & 1597.5 &  1.27 &           1p ($-16.6$) &                     np \\
  $\alpha$ Ori &     & 
      N6 & 1205.5 &   --- &           1p ($+24.3$) &           1p ($+23.3$) \\
         U Ori & 257 & 
      N4 & 1093.5 &  0.01 &           1p ($-23.7$) &           1p ($-23.7$) \\
 & & N17 & 1529.5 &  1.11 &                     np &           1p ($-22.4$) \\
 & & N18 & 1554.5 &  1.18 &                     np &           1p ($-23.3$) \\
 & & N20 & 1598.5 &  1.30 &                     np &           1p ($-23.4$) \\
\noalign{\smallskip}
\hline
\end{tabular}
\end{flushleft}}
\label{Tab:ccf_continued}
\end{table*}
      
\addtocounter{table}{-1}
\renewcommand{\baselinestretch}{0.7}
\begin{table*}
\caption[]{Continued}
\small{
\begin{flushleft}
\begin{tabular}{lllllll}
\hline\noalign{\smallskip}
Name & $\rho$      & Night  & Julian Date & Phase & 
K0-template & M4-template \\
     & ($R_\odot$) & Number & (2450000+)  &       & 
            &             \\
\noalign{\smallskip}
\hline\noalign{\smallskip}
  $\delta$ Aur &     & 
      N6 & 1205.5 &   --- &            1p ($+9.0$) &            1p ($+9.0$) \\
         X Aur & 125 & 
     N20 & 1598.5 &  0.42 &           1p ($-23.6$) &           1p ($-19.1$) \\
 & & N23 & 1683.5 &  0.94 &           1p ($-16.7$) &           1p ($-16.1$) \\
     $\mu$ Gem &     & 
      N6 & 1205.5 &   --- &           1p ($+54.5$) &           1p ($+54.4$) \\
        SU Cam & 198 & 
      N4 & 1093.5 &  0.03 &                    2p? &           1p ($-58.0$) \\
 & &  N6 & 1205.5 &  0.32 &                     np &           1p ($-57.5$) \\
 & &  N7 & 1233.5 &  0.41 &                     np &   2p ($-67.3$/$-56.6$) \\
 & & N17 & 1529.5 &  1.45 &                     np &           1p ($-61.9$) \\
 & & N18 & 1554.5 &  1.54 &                     np &           1p ($-63.7$) \\
 & & N19 & 1597.5 &  1.69 &                     np &           1p ($-62.6$) \\
 & & N22 & 1652.5 &  1.88 &           1p ($-61.6$) &           1p ($-60.9$) \\
 & & N23 & 1683.5 &  1.99 &           1p ($-63.3$) &           1p ($-60.9$) \\
         X Gem & 189 & 
     N20 & 1598.5 &  0.60 &                     np &           1p ($+39.9$) \\
 & & N22 & 1652.5 &  0.81 &           1p ($+42.0$) &           1p ($+42.6$) \\
 & & N23 & 1683.5 &  0.92 &           1p ($+40.7$) &           1p ($+42.5$) \\
         X Mon &     & 
     N22 & 1652.5 &  0.94 & ap ($+154.7$/$+171.5$) &          1p ($+167.0$) \\
         R Gem & 194 & 
      N4 & 1093.5 &  0.99 &   2p ($-56.6$/$-38.8$) &   2p ($-58.2$/$-39.2$) \\
 & &  N6 & 1205.5 &  1.21 &   2p ($-54.9$/$-40.5$) &   2p ($-53.0$/$-39.2$) \\
 & &  N7 & 1233.5 &  1.29 &   2p ($-55.0$/$-43.0$) &                     np \\
 & & N17 & 1529.5 &  2.09 &   2p ($-55.8$/$-39.2$) &                    2p? \\
 & & N18 & 1554.5 &  2.15 &   2p ($-55.6$/$-40.5$) &                     np \\
 & & N19 & 1597.5 &  2.27 &   2p ($-55.8$/$-41.7$) &                    1p? \\
 & & N22 & 1652.5 &  2.42 &                    1p? &                    1p? \\
         R CMi &     & 
     N20 & 1598.5 &  0.58 &           1p ($+45.9$) &                    1p? \\
         S CMi & 241 & 
      N6 & 1205.5 &  0.02 &                     np &           1p ($+67.5$) \\
 & &  N7 & 1233.5 &  0.10 &           1p ($+67.6$) &           1p ($+68.7$) \\
 & & N17 & 1529.5 &  0.99 &           1p ($+70.9$) &           1p ($+71.5$) \\
 & & N18 & 1554.5 &  1.07 &   2p ($+49.6$/$+72.5$) &           1p ($+72.9$) \\
 & & N19 & 1597.5 &  1.20 &                    2p? &           1p ($+73.3$) \\
$\upsilon$ Gem &     & 
      N6 & 1205.5 &   --- &           1p ($-21.7$) &           1p ($-21.6$) \\
        81 Gem &     & 
      N6 & 1205.5 &   --- &           1p ($+77.3$) &           1p ($+77.4$) \\
         R Cnc & 254 & 
      N3 & 1085.5 &  0.96 &                     np &           1p ($+28.0$) \\
 & & N17 & 1529.5 &  2.19 &                    1p? &           1p ($+31.7$) \\
 & & N18 & 1554.5 &  2.26 &                    1p? &           1p ($+32.1$) \\
 & & N20 & 1598.5 &  2.38 &                     np &           1p ($+30.4$) \\
 & & N22 & 1652.5 &  2.53 &                     np &   ap ($+19.5$/$+29.4$) \\
         X UMa & 161 & 
      N6 & 1205.5 &  0.05 &   2p ($-90.8$/$-77.0$) &           1p ($-79.0$) \\
 & &  N7 & 1233.5 &  0.16 &                    2p? &           1p ($-79.1$) \\
 & & N17 & 1529.5 &  1.35 &                     np &                    1p? \\
 & & N22 & 1652.5 &  1.85 &                     np &           1p ($-80.4$) \\
      HD 76830 &     & 
      N6 & 1205.5 &   --- &           1p ($+23.2$) &           1p ($+23.1$) \\
        UZ Hya & 173 & 
      N6 & 1205.5 &  0.98 &   ap ($+30.1$/$+46.3$) &           1p ($+40.4$) \\
 & &  N7 & 1233.5 &  1.09 &   ap ($+28.6$/$+44.8$) &   2p ($+27.4$/$+41.0$) \\
 & & N17 & 1529.5 &  2.22 &                    2p? &           1p ($+40.3$) \\
 & & N18 & 1554.5 &  2.32 &                     np &   2p ($+28.5$/$+41.1$) \\
 & & N20 & 1598.5 &  2.49 &                     np &           1p ($+34.1$) \\
         R Leo & 231 & 
     N19 & 1597.5 &  0.76 &                     np &            1p ($+5.2$) \\
 & & N22 & 1652.5 &  0.94 &                     np &            1p ($+7.1$) \\
 & & N23 & 1683.5 &  1.04 &                     np &            1p ($+7.5$) \\
         S LMi & 148 & 
     N20 & 1598.5 &  0.99 &    2p ($-14.8$/$+0.1$) &    ap ($-13.4$/$-6.2$) \\
 & & N22 & 1652.5 &  1.22 &           1p ($-12.1$) &            1p ($-9.2$) \\
         V Leo & 193 & 
      N6 & 1205.5 &  0.05 &   ap ($-26.5$/$-16.4$) &           1p ($-22.1$) \\
 & &  N7 & 1233.5 &  0.16 &                     np &           1p ($-21.8$) \\
 & & N17 & 1529.5 &  1.24 &                     np &           1p ($-21.4$) \\
 & & N18 & 1554.5 &  1.33 &                     np &           1p ($-19.8$) \\
         R UMa & 181 & 
      N9 & 1291.5 &  0.03 &   2p ($+14.2$/$+36.5$) &           1p ($+35.1$) \\
 & & N17 & 1529.5 &  0.82 &                     np &           1p ($+33.7$) \\
 & & N18 & 1554.5 &  0.90 &                    1p? &           1p ($+34.1$) \\
 & & N19 & 1597.5 &  1.04 &           1p ($+35.2$) &           1p ($+33.8$) \\
        RU UMa & 162 & 
      N6 & 1205.5 &  0.97 &                     np &           1p ($-56.2$) \\
 & &  N7 & 1233.5 &  1.08 &   2p ($-67.5$/$-54.8$) &           1p ($-56.1$) \\
 & & N17 & 1529.5 &  2.26 &                     np &           1p ($-57.6$) \\
 & & N18 & 1554.5 &  2.35 &                     np &                     np \\
         Y Vir & 142 & 
      N7 & 1233.5 &  0.99 &    2p ($+3.4$/$+16.6$) &           1p ($+13.3$) \\
 & & N18 & 1554.5 &  2.46 &                     np &                    2p? \\
 & & N20 & 1598.5 &  2.66 &                     np &            1p ($+7.3$) \\
 & & N22 & 1652.5 &  2.90 &    2p ($+0.4$/$+14.4$) &            1p ($+9.1$) \\
         R Vir & 117 & 
     N23 & 1683.5 &  0.92 &   ap ($-35.3$/$-23.3$) &           1p ($-32.8$) \\
 & & N24 & 1716.5 &  1.14 &   ap ($-35.1$/$-23.0$) &           1p ($-31.0$) \\
\noalign{\smallskip}
\hline
\end{tabular}
\end{flushleft}}
\label{Tab:ccf_continued}
\end{table*}
      
\addtocounter{table}{-1}
\renewcommand{\baselinestretch}{0.7}
\begin{table*}
\caption[]{Continued}
\small{
\begin{flushleft}
\begin{tabular}{lllllll}
\hline\noalign{\smallskip}
Name & $\rho$      & Night  & Julian Date & Phase & 
K0-template & M4-template \\
     & ($R_\odot$) & Number & (2450000+)  &       & 
            &             \\
\noalign{\smallskip}
\hline\noalign{\smallskip}
        RS UMa & 172 & 
      N6 & 1205.5 &  0.10 &   ap ($-31.8$/$-18.1$) &           1p ($-30.3$) \\
 & &  N7 & 1233.5 &  0.20 &   2p ($-37.3$/$-23.8$) &           1p ($-29.9$) \\
 & & N18 & 1554.5 &  1.44 &                     np &           1p ($-25.8$) \\
 & & N24 & 1716.5 &  2.07 &           1p ($-23.3$) &           1p ($-23.6$) \\
         S UMa & 133 & 
     N19 & 1597.5 &  0.91 &    ap ($-1.0$/$+14.0$) &    ap ($-1.8$/$+11.6$) \\
 & & N22 & 1652.5 &  1.15 &            1p ($-7.7$) &            1p ($-7.5$) \\
         U Vir & 137 & 
     N20 & 1598.5 &  0.06 &   2p ($-56.3$/$-41.7$) &   2p ($-56.3$/$-44.5$) \\
 & & N22 & 1652.5 &  0.32 &                     np &           1p ($-47.8$) \\
         V UMi &     & 
      N7 & 1233.5 & 0.99: &          1p ($-171.2$) &          1p ($-171.9$) \\
 & & N18 & 1554.5 & 5.44: &          1p ($-165.9$) &          1p ($-167.0$) \\
 & & N19 & 1597.5 & 6.04: &          1p ($-170.3$) &          1p ($-171.8$) \\
 & & N22 & 1652.5 & 6.81: &          1p ($-168.7$) &          1p ($-166.7$) \\
 & & N24 & 1716.5 & 7.69: &          1p ($-168.3$) &          1p ($-166.0$) \\
        SY Vir & 197 & 
     N24 & 1716.5 &  0.05 &   ap ($+11.2$/$+21.7$) &           1p ($+17.9$) \\
     HD 123657 &     & 
      N6 & 1205.5 &   --- &           1p ($-37.3$) &           1p ($-37.5$) \\
  $\alpha$ Boo &     & 
     N19 & 1597.5 &   --- &            1p ($-5.3$) &            1p ($-5.2$) \\
         R Boo & 151 & 
      N7 & 1233.5 &  0.98 &   2p ($-66.4$/$-49.1$) &   ap ($-62.8$/$-51.9$) \\
 & &  N9 & 1291.5 &  1.24 &                     np &           1p ($-55.1$) \\
 & & N18 & 1554.5 &  2.41 &                     np &   ap ($-64.1$/$-53.8$) \\
 & & N19 & 1597.5 &  2.61 &                     np &           1p ($-59.7$) \\
 & & N22 & 1652.5 &  2.85 &           1p ($-55.2$) &           1p ($-56.5$) \\
         Y Lib & 194 & 
      N6 & 1205.5 &  0.98 &                     np &     2p ($-2.9$/$+7.1$) \\
 & &  N7 & 1233.5 &  1.08 &                    1p? &            1p ($+3.1$) \\
 & & N19 & 1597.5 &  2.40 &                     np &            1p ($+6.4$) \\
 & & N24 & 1716.5 &  2.83 &                     np &            1p ($+4.3$) \\
        RT Boo & 215 & 
     N20 & 1598.5 &  0.05 &           1p ($+34.2$) &           1p ($+37.3$) \\
 & & N23 & 1683.5 &  0.36 &                     np &           1p ($+34.4$) \\
         S Ser & 232 & 
     N25 & 1737.5 &  0.01 &                     np &            1p ($+9.2$) \\
         S UMi & 241 & 
      N1 & 1031.5 &  0.69 &                     np &           1p ($-53.8$) \\
 & &  N6 & 1205.5 &  1.22 &                     np &                     np \\
 & &  N7 & 1233.5 &  1.30 &                     np &   ap ($-56.5$/$-48.3$) \\
 & &  N9 & 1291.5 &  1.48 &                     np &           1p ($-54.7$) \\
 & & N12 & 1365.5 &  1.70 &                     np &           1p ($-52.4$) \\
 & & N13 & 1386.5 &  1.76 &                     np &           1p ($-51.5$) \\
 & & N14 & 1424.5 &  1.88 &                     np &           1p ($-50.0$) \\
 & & N20 & 1598.5 &  2.40 &                     np &   ap ($-59.2$/$-51.0$) \\
 & & N22 & 1652.5 &  2.57 &                     np &           1p ($-55.6$) \\
 & & N24 & 1716.5 &  2.76 &                    1p? &           1p ($-52.8$) \\
        ST Her &     & 
      N7 & 1233.5 &   --- &           1p ($-24.3$) &           1p ($-22.8$) \\
        RU Her & 303 & 
      N9 & 1291.5 &  0.02 &           1p ($-25.0$) &           1p ($-25.2$) \\
 & & N10 & 1318.5 &  0.08 &           1p ($-26.0$) &           1p ($-25.1$) \\
 & & N12 & 1365.5 &  0.17 &                     np &           1p ($-25.0$) \\
 & & N20 & 1598.5 &  0.65 &                     np &           1p ($-29.9$) \\
 & & N23 & 1683.5 &  0.83 &                     np &           1p ($-28.3$) \\
 & & N24 & 1716.5 &  0.90 &                     np &           1p ($-28.0$) \\
 & & N26 & 1766.5 &  1.00 &                    1p? &           1p ($-27.1$) \\
        SS Oph & 150 & 
     N24 & 1716.5 &  0.98 &   2p ($-41.9$/$-28.2$) &           1p ($-31.5$) \\
        RV Her & 137 & 
     N24 & 1716.5 &  0.95 &                    2p? &           1p ($-47.9$) \\
        SY Her &  93 & 
     N23 & 1683.5 &  0.09 &   ap ($+21.6$/$+36.6$) &           1p ($+31.2$) \\
         Z Oph & 143 & 
      N1 & 1031.5 &  0.08 &   2p ($-96.0$/$-77.2$) &   2p ($-97.1$/$-75.7$) \\
 & &  N2 & 1060.5 &  0.17 &   2p ($-93.7$/$-78.5$) &   2p ($-95.2$/$-79.4$) \\
 & &  N9 & 1291.5 &  0.83 &                    1p? &           1p ($-83.2$) \\
 & & N10 & 1318.5 &  0.91 &                     np &                     np \\
 & & N12 & 1365.5 &  1.04 &   2p ($-97.8$/$-76.1$) &           1p ($-79.7$) \\
 & & N13 & 1386.5 &  1.10 &   2p ($-94.6$/$-76.1$) &           1p ($-79.4$) \\
 & & N14 & 1424.5 &  1.21 &   2p ($-88.2$/$-75.5$) &           1p ($-79.0$) \\
 & & N22 & 1652.5 &  1.86 &           1p ($-73.9$) &           1p ($-75.6$) \\
 & & N24 & 1716.5 &  2.05 &   2p ($-94.3$/$-76.9$) &           1p ($-74.1$) \\
 & & N27 & 1767.5 &  2.19 &   2p ($-91.6$/$-75.0$) &   2p ($-94.8$/$-76.0$) \\
        RS Her & 156 & 
      N9 & 1291.5 &  0.95 &           1p ($-34.7$) &           1p ($-33.8$) \\
 & & N10 & 1318.5 &  1.07 &           1p ($-40.3$) &           1p ($-34.1$) \\
 & & N25 & 1737.5 &  2.98 &   2p ($-44.0$/$-26.8$) &   ap ($-44.5$/$-30.4$) \\
        RU Oph & 142 & 
     N24 & 1716.5 &  0.00 &           1p ($+42.0$) &           1p ($+40.5$) \\
   $\beta$ Oph &     & 
      N9 & 1291.5 &   --- &           1p ($-12.2$) &           1p ($-12.2$) \\
 & & N10 & 1318.5 &   --- &           1p ($-12.5$) &           1p ($-12.4$) \\
 & & N12 & 1365.5 &   --- &           1p ($-12.6$) &           1p ($-12.6$) \\
 & & N13 & 1386.5 &   --- &           1p ($-12.4$) &           1p ($-12.4$) \\
 & & N23 & 1683.5 &   --- &           1p ($-12.3$) &           1p ($-12.3$) \\
 & & N26 & 1766.5 &   --- &           1p ($-12.4$) &           1p ($-12.7$) \\
\noalign{\smallskip}
\hline
\end{tabular}
\end{flushleft}}
\label{Tab:ccf_continued}
\end{table*}
      
\addtocounter{table}{-1}
\renewcommand{\baselinestretch}{0.7}
\begin{table*}
\caption[]{Continued}
\small{
\begin{flushleft}
\begin{tabular}{lllllll}
\hline\noalign{\smallskip}
Name & $\rho$      & Night  & Julian Date & Phase & 
K0-template & M4-template \\
     & ($R_\odot$) & Number & (2450000+)  &       & 
            &             \\
\noalign{\smallskip}
\hline\noalign{\smallskip}
         T Her & 120 & 
     N24 & 1716.5 &  0.03 & 2p ($-133.4$/$-115.0$) &          1p ($-117.9$) \\
        RY Oph & 119 & 
     N22 & 1652.5 &  0.95 &   2p ($-75.9$/$-54.7$) &           1p ($-56.6$) \\
        RT Dra & 195 & 
      N2 & 1060.5 &  0.04 &                    2p? &           1p ($-99.8$) \\
 & &  N3 & 1085.5 &  0.13 &                     np &          1p ($-101.5$) \\
 & &  N4 & 1093.5 &  0.27 &                     np & ap ($-111.8$/$-100.6$) \\
 & &  N5 & 1171.5 &  0.44 &                     np &                     np \\
 & &  N9 & 1291.5 &  0.87 &          1p ($-106.3$) &          1p ($-104.4$) \\
 & & N10 & 1318.5 &  0.96 &          1p ($-108.8$) &          1p ($-103.8$) \\
 & & N12 & 1365.5 &  1.13 &                     np &          1p ($-104.0$) \\
 & & N13 & 1386.5 &  1.21 &                     np &          1p ($-103.7$) \\
 & & N14 & 1424.5 &  1.34 &                     np &          1p ($-105.5$) \\
 & & N15 & 1450.5 &  1.44 &                     np &          1p ($-106.2$) \\
 & & N22 & 1652.5 &  2.16 &                     np &          1p ($-100.0$) \\
 & & N23 & 1683.5 &  2.27 &                     np & 2p ($-112.6$/$-101.9$) \\
 & & N24 & 1716.5 &  2.39 &                     np &          1p ($-109.4$) \\
 & & N26 & 1766.5 &  2.57 &                     np &          1p ($-107.4$) \\
        SV Her & 178 & 
     N27 & 1767.5 &  0.94 &           1p ($-13.0$) &           1p ($-13.9$) \\
         X Oph & 216 & 
     N13 & 1386.5 &  0.90 &           1p ($-73.2$) &           1p ($-71.4$) \\
 & & N14 & 1424.5 &  1.02 &           1p ($-72.8$) &           1p ($-70.8$) \\
 & & N15 & 1450.5 &  1.09 &           1p ($-73.1$) &           1p ($-70.1$) \\
 & & N26 & 1766.5 &  2.06 &           1p ($-74.1$) &           1p ($-75.2$) \\
        WZ Lyr & 429 & 
     N24 & 1716.5 &  0.97 &                    1p? &            1p ($+1.7$) \\
 & & N26 & 1766.5 &  1.10 &                    1p? &            1p ($+1.4$) \\
        RU Lyr & 258 & 
     N27 & 1767.5 &  0.98 &                    1p? &            1p ($-3.3$) \\
         W Aql & 242 & 
     N23 & 1683.5 &  0.97 &   2p ($-53.5$/$-27.4$) &           1p ($-23.4$) \\
        RT Cyg & 130 & 
      N1 & 1031.5 &  0.95 & 2p ($-132.5$/$-110.1$) &          1p ($-111.4$) \\
 & &  N2 & 1060.5 &  1.10 & 2p ($-130.7$/$-111.0$) & 2p ($-134.9$/$-111.6$) \\
 & &  N3 & 1085.5 &  1.23 & ap ($-122.5$/$-110.3$) &          1p ($-112.5$) \\
 & & N12 & 1365.5 &  2.70 &                     np &                     np \\
 & & N13 & 1386.5 &  2.81 &          1p ($-113.0$) &          1p ($-114.6$) \\
 & & N14 & 1424.5 &  3.01 & 2p ($-130.2$/$-110.4$) &          1p ($-112.0$) \\
 & & N15 & 1450.5 &  3.15 & 2p ($-128.5$/$-111.1$) & 2p ($-130.7$/$-112.7$) \\
  $\gamma$ Aql &     & 
      N9 & 1291.5 &   --- &            1p ($-3.1$) &            1p ($-2.8$) \\
 & & N23 & 1683.5 &   --- &            1p ($-3.1$) &            1p ($-2.9$) \\
 & & N26 & 1766.5 &   --- &            1p ($-3.2$) &            1p ($-3.3$) \\
    $\chi$ Cyg & 273 & 
      N1 & 1031.5 &  0.76 &                     np &            1p ($-7.1$) \\
 & &  N2 & 1060.5 &  0.83 &                     np &            1p ($-5.5$) \\
 & &  N3 & 1085.5 &  0.89 &            1p ($-5.1$) &            1p ($-5.3$) \\
 & &  N4 & 1093.5 &  0.99 &            1p ($-5.1$) &            1p ($-5.7$) \\
 & & N14 & 1424.5 &  1.73 &                     np &            1p ($-5.5$) \\
 & & N15 & 1450.5 &  1.79 &                     np &            1p ($-5.0$) \\
         Z Cyg & 189 & 
     N26 & 1766.5 &  0.03 & 2p ($-179.4$/$-160.5$) &          1p ($-160.9$) \\
         S Aql &     & 
     N27 & 1767.5 &  0.95 & ap ($-114.7$/$-100.6$) &          1p ($-103.2$) \\
         Z Aql & 109 & 
     N23 & 1683.5 &  0.08 &    2p ($-16.5$/$+2.4$) &            1p ($-3.3$) \\
        WX Cyg &     & 
      N3 & 1085.5 &  0.96 &   2p ($+30.6$/$+47.6$) &                     np \\
 & &  N4 & 1093.5 &  1.06 &   2p ($+25.8$/$+45.4$) &                     np \\
         T Cep & 238 & 
      N1 & 1031.5 &  0.23 &                    2p? &           1p ($-10.9$) \\
 & &  N2 & 1060.5 &  0.31 &                     np &           1p ($-10.9$) \\
 & & N12 & 1365.5 &  1.09 &                    1p? &           1p ($-11.3$) \\
 & & N14 & 1424.5 &  1.25 &                     np &           1p ($-11.1$) \\
 & & N23 & 1683.5 &  1.91 &                     np &           1p ($-14.9$) \\
 & & N27 & 1767.5 &  2.13 &           1p ($-15.8$) &           1p ($-13.9$) \\
        RR Aqr & 127 & 
     N27 & 1767.5 &  0.04 & 2p ($-191.0$/$-174.4$) & 2p ($-191.3$/$-178.2$) \\
         X Peg & 135 & 
     N23 & 1683.5 &  0.01 &   ap ($-49.1$/$-35.5$) &           1p ($-39.7$) \\
        SW Peg & 222 & 
      N3 & 1085.5 &  0.96 &            1p ($-3.8$) &            1p ($-4.4$) \\
 & & N15 & 1450.5 &  1.88 &                     np &            1p ($-4.2$) \\
        AX Cep &     & 
      N2 & 1060.5 & 0.19: &     ap ($-5.3$/$+7.5$) &                     np \\
 & &  N3 & 1085.5 & 0.25: &            1p ($-2.1$) &                     np \\
 & &  N4 & 1093.5 & 0.35: &            1p ($-1.0$) &                     np \\
 & &  N5 & 1171.5 & 0.47: &            1p ($+1.6$) &                     np \\
 & &  N7 & 1233.5 & 0.63: &            1p ($+6.7$) &                     np \\
 & & N13 & 1386.5 & 1.01: &     2p ($-7.2$/$+7.6$) &                     np \\
 & & N14 & 1424.5 & 1.11: &     ap ($-6.4$/$+6.3$) &                     np \\
 & & N15 & 1450.5 & 1.17: &            1p ($-3.7$) &                     np \\
 & & N17 & 1529.5 & 1.37: &            1p ($-1.6$) &                     np \\
 & & N18 & 1554.5 & 1.44: &            1p ($+0.0$) &                     np \\
 & & N26 & 1766.5 & 1.97: &     2p ($-9.9$/$+7.6$) &                     np \\
         S Cep &     & 
      N1 & 1031.5 &  0.89 &   2p ($-38.5$/$-18.2$) &                     np \\
 & &  N2 & 1060.5 &  0.95 &   2p ($-37.2$/$-18.5$) &                     np \\
\noalign{\smallskip}
\hline
\end{tabular}
\end{flushleft}}
\label{Tab:ccf_continued}
\end{table*}
      
\addtocounter{table}{-1}
\renewcommand{\baselinestretch}{0.7}
\begin{table*}
\caption[]{Continued}
\small{
\begin{flushleft}
\begin{tabular}{lllllll}
\hline\noalign{\smallskip}
Name & $\rho$      & Night  & Julian Date & Phase & 
K0-template & M4-template \\
     & ($R_\odot$) & Number & (2450000+)  &       & 
            &             \\
\noalign{\smallskip}
\hline\noalign{\smallskip}
 & &  N3 & 1085.5 &  1.00 &   2p ($-36.0$/$-18.7$) &                     np \\
 & &  N4 & 1093.5 &  1.08 &   2p ($-33.7$/$-18.6$) &                     np \\
 & &  N7 & 1233.5 &  1.30 &           1p ($-27.0$) &                     np \\
 & & N12 & 1365.5 &  1.58 &           1p ($-22.0$) &                     np \\
 & & N13 & 1386.5 &  1.62 &           1p ($-21.1$) &                     np \\
 & & N14 & 1424.5 &  1.70 &           1p ($-20.7$) &                     np \\
 & & N15 & 1450.5 &  1.75 &           1p ($-20.7$) &                     np \\
 & & N17 & 1529.5 &  1.91 &   2p ($-39.0$/$-20.0$) &                     np \\
 & & N18 & 1554.5 &  1.96 &   2p ($-39.0$/$-20.4$) &                     np \\
 & & N26 & 1766.5 &  2.40 &           1p ($-26.3$) &                     np \\
        RV Peg & 268 & 
     N27 & 1767.5 &  0.97 &           1p ($-32.4$) &           1p ($-32.5$) \\
        AR Cep &     & 
      N1 & 1031.5 &   --- &           1p ($-16.5$) &           1p ($-16.3$) \\
 & &  N4 & 1093.5 &   --- &           1p ($-16.0$) &           1p ($-16.1$) \\
         R Peg & 261 & 
      N2 & 1060.5 &  0.97 &           1p ($+20.7$) &           1p ($+21.7$) \\
 & &  N3 & 1085.5 &  1.03 &           1p ($+20.2$) &           1p ($+21.9$) \\
 & & N15 & 1450.5 &  2.00 &                    1p? &           1p ($+21.5$) \\
 & & N17 & 1529.5 &  2.21 &                     np &           1p ($+21.8$) \\
         W Peg & 247 & 
      N2 & 1060.5 &  0.02 &                    1p? &           1p ($-19.5$) \\
 & &  N3 & 1085.5 &  0.10 &                    2p? &           1p ($-18.4$) \\
 & & N13 & 1386.5 &  0.97 &                     np &           1p ($-25.0$) \\
 & & N15 & 1450.5 &  1.15 &                     np &           1p ($-22.5$) \\
 & & N27 & 1767.5 &  2.07 &                    2p? &           1p ($-18.1$) \\
         S Peg & 212 & 
      N2 & 1060.5 &  0.98 &                    1p? &            1p ($+5.3$) \\
 & &  N3 & 1085.5 &  1.06 &                    1p? &            1p ($+6.0$) \\
 & & N13 & 1386.5 &  2.00 &                    1p? &            1p ($+4.8$) \\
        RY Cep &  86 & 
      N2 & 1060.5 & 0.60: &          1p ($-231.7$) &          1p ($-233.9$) \\
 & &  N3 & 1085.5 & 0.77: &          1p ($-235.1$) &          1p ($-230.8$) \\
 & &  N4 & 1093.5 & 1.03: &          1p ($-245.1$) & 2p ($-249.5$/$-229.3$) \\
 & &  N7 & 1233.5 & 1.76: &          1p ($-234.2$) &          1p ($-234.3$) \\
 & & N13 & 1386.5 & 2.79: & ap ($-234.9$/$-222.2$) &          1p ($-235.6$) \\
 & & N15 & 1450.5 & 3.22: &          1p ($-241.5$) & ap ($-244.5$/$-231.9$) \\
 & & N17 & 1529.5 & 3.75: &          1p ($-230.8$) &          1p ($-232.6$) \\
 & & N18 & 1554.5 & 3.92: & 2p ($-242.9$/$-226.6$) &          1p ($-229.6$) \\
 & & N27 & 1767.5 & 5.35: &          1p ($-239.2$) & 2p ($-246.1$/$-234.0$) \\
        ST And &     & 
      N4 & 1093.5 &  0.99 &   2p ($+27.5$/$+43.1$) &                     np \\
 & & N15 & 1450.5 &  1.98 &   ap ($+28.7$/$+43.5$) &                     np \\
 & & N17 & 1529.5 &  2.22 &   ap ($+25.0$/$+33.8$) &                     np \\
 & & N18 & 1554.5 &  2.30 &   2p ($+27.1$/$+35.9$) &                     np \\
 & & N26 & 1766.5 &  2.95 &           1p ($+37.2$) &                     np \\
         R Cas & 282 & 
      N4 & 1093.5 &  0.41 &                     np &                    1p? \\
 & & N13 & 1386.5 &  1.02 &                    1p? &           1p ($+22.7$) \\
 & & N17 & 1529.5 &  1.35 &                     np &           1p ($+20.1$) \\
 & & N26 & 1766.5 &  1.90 &           1p ($+21.3$) &           1p ($+21.4$) \\
\noalign{\smallskip}
\hline
\end{tabular}
\end{flushleft}}
\label{Tab:ccf}
\end{table*}

\end{document}